\documentclass[prb,twocolumn,noshowkeys,preprintnumbers,amsmath,%
amssymb,showpacs,aps,10pt,superscriptaddress,floatx]{revtex4}
\usepackage{graphicx,color}

\usepackage{amsmath,enumerate}
\usepackage{bbm,amsfonts}

\newcommand{\be}{\begin{equation}}
\newcommand{\ee}{\end{equation}}
\newcommand{\bea}{\begin{eqnarray}}
\newcommand{\eea}{\end{eqnarray}}

\newcommand{\hideit}[1]{}

\begin{document}

\title{Topological and Entanglement Properties of Resonating Valence Bond wavefunctions}

\author{Didier \surname{Poilblanc} }
\affiliation{Laboratoire de Physique Th\'eorique, C.N.R.S. and Universit\'e de Toulouse, 31062 Toulouse, France}
\author{Norbert Schuch}
\affiliation{Institute for Quantum Information, California Institute of
Technology, MC 305-16, Pasadena CA 91125, U.S.A.}
\affiliation{Institut f\"ur Quanteninformation, RWTH Aachen, D-52056 Aachen, Germany}
\author{David P\'erez-Garc\'{\i}a}
\affiliation{Department of Mathematical Analysis, Faculty of Mathematics, UCM, Spain}
\author{J.~Ignacio \surname{Cirac}}
\affiliation{Max-Planck-Institut f{\"{u}}r Quantenoptik,
Hans-Kopfermann-Str.\ 1, D-85748 Garching, Germany}

\begin{abstract}
We examine in details the connections between topological and entanglement properties of 
short-range resonating valence bond (RVB) wave functions using Projected Entangled Pair States
(PEPS) on kagome and square lattices on (quasi-)infinite cylinders with generalized boundary conditions (and perimeters with up to 20 lattice spacings).
Making use of disconnected topological sectors in the space 
of dimer lattice coverings, we explicitly derive (orthogonal) ``minimally entangled"
PEPS RVB states. For the kagome lattice,  using the quantum Heisenberg antiferromagnet  
as a reference model, we obtain the finite size scaling with increasing cylinder perimeter of the vanishing energy separations 
between these states. In particular, we extract two separate (vanishing) energy scales corresponding
(i) to insert a vison line between the two ends of the cylinder and (ii) to pull out and 
freeze a spin at either end.
We also investigate the relations between bulk and boundary properties and show that, for a bipartition
of the cylinder, 
the boundary Hamiltonian defined on the edge can be written as a product of a highly non-local 
projector, which fundamentally depends 
upon boundary conditions, with an emergent (local)  $su(2)$-invariant
one-dimensional (superfluid) t--J Hamiltonian, which arises due to the symmetry 
properties of the auxiliary spins at the edge. 
This multiplicative structure, a consequence of the disconnected topological sectors in the space 
of dimer lattice coverings, is characteristic of the topological nature of the states.
For minimally entangled RVB states, it is shown that the entanglement spectrum, which
reflects the properties of the (gapless or gapped) edge modes, is a subset of the spectrum of 
the local Hamiltonian, e.g. half of it for the kagome RVB state, providing a simple argument on 
the origin of the topological entanglement entropy $S_0= -\ln{2}$ of the $\mathbb{Z}_2$ spin liquid.
We propose to use these features to probe topological phases in microscopic Hamiltonians
and some results are compared to existing DMRG data. 
\end{abstract}


\maketitle

\section{Introduction}

Conventional ordering in condensed matter systems is traditionally associated to
symmetry-breaking and to the existence of a local order 
parameter (Landau theory).
Topologically ordered phases of matter~\cite{wen-91} offer
completely new classes of systems for which the ground state (GS) degeneracy depends 
on topology (disc, cylinder, torus, etc...). The (short-range singlet) Resonating Valence Bond 
(RVB) wavefunction proposed by Anderson~\cite{anderson}
as the parent Mott insulator of high-temperature superconductors is a celebrated example. 
Such topological 
phases carry emerging fractionalized excitations and  raise growing attention 
due to their potential to realize
fault-tolerant setups for quantum computing.~\cite{kitaev:toriccode}

Experimental and theoretical search for topological liquids in quantum 
antiferromagnets~\cite{Mendels}
and in related microscopic models~\cite{review,Assaad} has been a long-standing quest. 
One major problem is the existence of many possible (non-magnetic) nearby 
competing states
like valence bond crystals~\cite{vmc} (spontaneously breaking lattice symmetry), 
clearly evidenced e.g. in quantum dimer models.~\cite{vbs}
Recent advances in the Density Matrix Renormalisation Group (DMRG) techniques
has reinforced the strong belief that a gapped spin liquid might be stabilized in the 
nearest neighbor (NN) S=1/2 Heisenberg quantum antiferromagnet (HAF).~\cite{White,Balents,Balents2,Schollwoeck}
This has also triggered the search for novel theoretical tools capable of better
detecting topological order, in particular entanglement measures used in quantum information.
A common setup consists of dividing the system into two regions (named A and B)
and compute the reduced density matrix (RDM) in the GS of e.g. the A subsystem.
The entanglement entropy (EE), defined as the Von Neumann entropy of the RDM 
$S_{\rm VN}=-\rho_A \ln{\rho_A}$, contains an extensive term -- proportional to
the length of the boundary (area law) -- and a universal sub-leading constant, the topological EE.
Specific disc-like setups~\cite{KP2006} or cylindrical geometries can be used
to extract the topological EE. 

In fact, $-\ln{\rho_A}$ can be seen as a (dimensionless) Hamiltonian $H_b$,
a key conceptual object. First, its spectrum, the so-called entanglement spectrum
(ES), has been conjectured to show a one-to-one correspondence with the spectrum of edge 
states. This remarkable property was first established in fractional quantum 
Hall states~\cite{entspectrum} and, then, in quantum spin systems~\cite{poilblanc2010}. 
Furthermore, 
Projected Engangled Pair States (PEPS)~\cite{PEPS} offer a natural
formulation of the relation between bulk and boundary.  In
Ref.~\onlinecite{ciracpoilblanc2011}, an explicit isometry was constructed
which maps the Hamiltonian $H_b$ onto another one $\tilde  H_b$ acting on
the
space of auxiliary spins living at the edge 
of region A, while keeping the spectrum. Furthermore, for various 
two-dimensional (2D) models displaying quantum phase transitions, like a deformed AKLT~\cite{AKLT} or an Ising-type~\cite{verstraetewolf06} model,
it was found~\cite{ciracpoilblanc2011} that a gapped bulk phase with
local order corresponds to a boundary Hamiltonian with local interactions, whereas critical behavior
in the bulk is reflected in a diverging interaction length of $\tilde H_b$. 

Entanglement properties of 2D topological phases are less well understood. 
Rokhsar-Kivelson (RK) wave functions, defined as equal-weight superposition of fully packed dimer coverings, exhibit critical behavior on bipartite lattices~\cite{RK0} or realize the simplest topological phase, 
the so-called $\mathbb{Z}_2$ liquid, on frustrated lattices.~\cite{RK1} 
The topological EE of critical and topological RK wavefunctions have been computed using various 
topologies~\cite{EE_RK} and
the boundary Hamiltonian corresponding to the GS 
of Kitaev's toric code~\cite{kitaev:toriccode} was shown 
to be {\it non-local}.~\cite{ciracpoilblanc2011}
Unfortunately, RK-like wavefunctions are not generic -- their ES 
is completely dispersionless~\cite{ES_RK} -- and do not 
describe real quantum $S=1/2$ spin systems.~\cite{note1,long_paper} 
In contrast, (short-range) RVB states, defined as linear superposition of 
hardcore coverings of {\it non-orthogonal} nearest-neighbor 
{\it SU(2) singlets} (see Fig.~\ref{Fig:cylinders}(a)), appear to be closer to 
physical systems. 
Very recently,  the (Renyi) EE between (finite) cylindrical regions has been computed~\cite{EE_2D_QMC} numerically for the critical~\cite{critical_QMC,note_CL}  RVB state on the square lattice.
Similarly, the (Renyi) topological EE of SU(2)-symmetric gapped chiral and $\mathbb{Z}_2$ spin liquids 
was obtained~\cite{zhanggrover} using Kitaev-Preskill prescription. 
Nevertheless, ES and boundary Hamiltonians of 
such RVB/spin liquids wavefunctions are unknown. 

In this work, we study topological and entanglement properties of both critical (square lattice) and
gapped topologically-ordered (kagome lattice) RVB wavefunctions~\cite{note_CL} on {\it infinite} cylinders making use of
simple PEPS representations. Let us describe here the organization of the paper:
First, in Sec.~\ref{Sec:rvb} we introduce RVB wavefunctions defined in the space 
of dimer (hardcore) coverings of square and kagome lattices.
On cylinders with generalized boundary conditions, we review the construction of four disconnected 
topological sectors of dimer coverings (on the kagome lattice). 
Next, in Sec.~\ref{Sec:peps} we introduce the PEPS
representation of the RVB wavefunctions and, making use of the disconnected topological sectors,
explicitly construct four orthogonal RVB states. Using the quantum Heisenberg model as
a reference Hamiltonian,  we obtain  the generic behavior of their energy splittings versus
cylinder perimeter. 
In Sec.~\ref{Sec:boundary}, we introduce a partition of the cylinder and compute the 
corresponding Reduced Density Matrix (RDM). 
The (hermitian) operator defined as minus the logarithm of the RDM can be viewed as a boundary Hamiltonian:
it is can be naturally expressed in the PEPS formalism
as an operator acting on the virtual indices on the edges (up to an isometry).
We show that the boundary Hamiltonian can be written
as a product of a highly non-local projector, which depends fundamentally on the boundary conditions,
by a local one-dimensional t--J model, which arises due to the symmetry properties of the auxiliary spins at the boundary and characterizes the (gapless or gapped) edge modes.
This multiplicative structure is a direct consequence of the disconnected topological sectors in the space 
of dimer coverings of the lattice and, therefore,
reflects the topological nature of the states. For sake of conciseness, more technical issues such as finite size scalings, etc... are treated in Appendices.  
 
\section{RVB wavefunctions on cylinders}
\label{Sec:rvb}

\subsection{Set-up and boundary conditions}

\begin{figure}
\includegraphics[width=0.95\columnwidth]{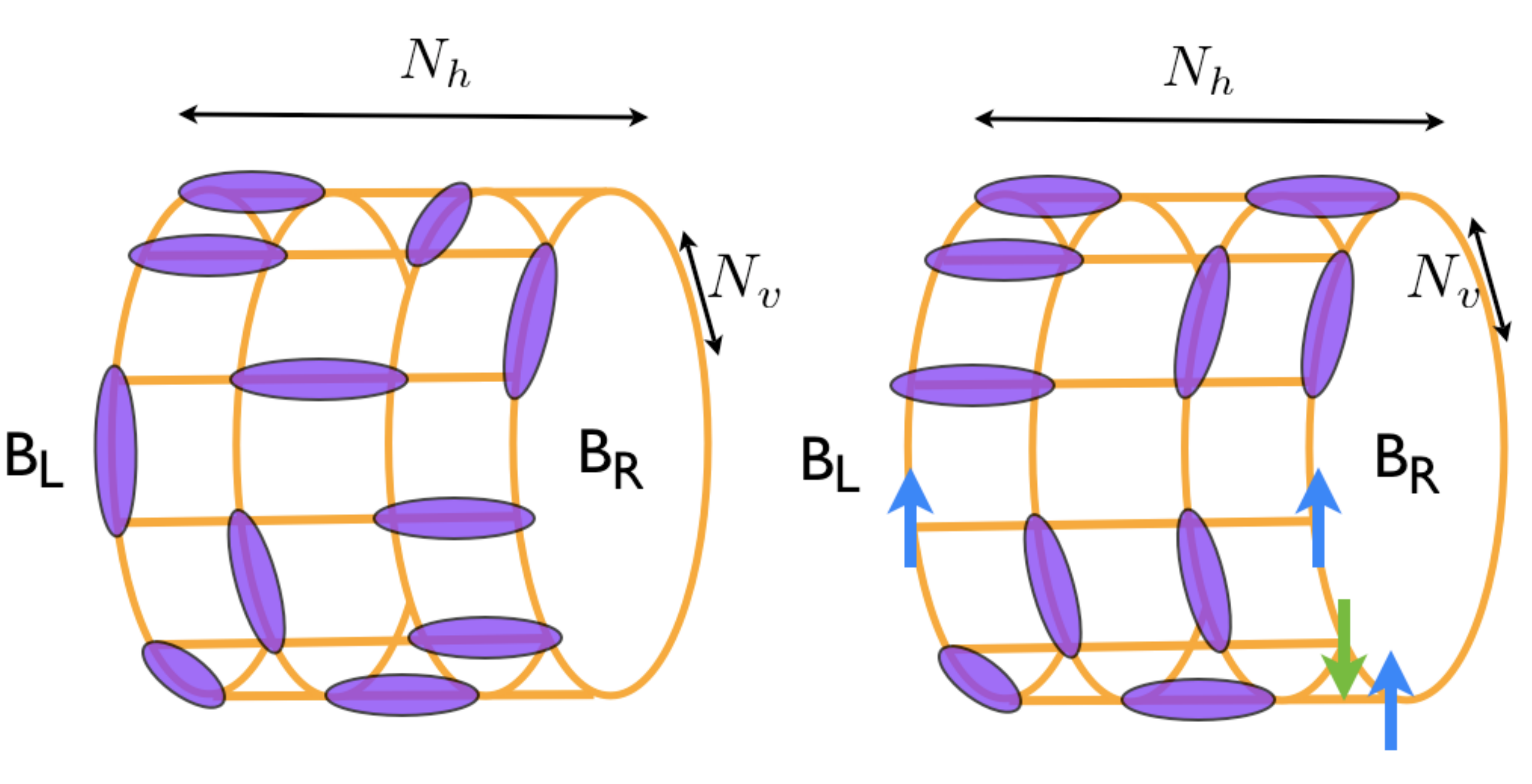}
\caption{(Color online) Typical valence bond configurations on a $N_v\times N_h$
cylinder with periodic boundary conditions
along the vertical ($v$) direction.
Ellipses represent
singlets of two spins 1/2. Open (a) or generalized (b) boundary conditions on the 
$B_L$ and $B_R$ ends of the cylinder are considered [GBC 
can be obtained physically by freezing some spins at the boundaries, e.g. with
local magnetic fields]. The RVB wavefunction is defined as the equal-weight 
superposition of all such configurations (for a fixed realization of $B_L$ and $B_R$).}
\label{Fig:cylinders}
\end{figure}

Let us first start with a square lattice on a cylinder of length $N_h$ and circumference $N_v$
with Open Boundary Conditions (OBC) as depicted in Fig.~\ref{Fig:cylinders}(a).
We consider the space of all nearest-neighbor  (NN)
\hbox{$|\uparrow\downarrow\rangle-|\downarrow\uparrow\rangle$} singlet
coverings of the lattice in such a way that each site belongs to one and only one dimer
(so called ``hard-core" coverings).
Note that all singlets are oriented from one sublattice to the other.
The Resonating Valence Bond state is then defined as the equal weight superposition of all such 
dimer (singlet) coverings. Besides OBC we also consider Generalized Boundary Conditions (GBC)
as in Fig.~\ref{Fig:cylinders}(b) by freezing some spins at the two boundaries 
$B_L$ and $B_R$ of the cylinder: in that case, dimers cannot involve these ``frozen" sites
any more. Because of the local hard-core constraints, the choice of the boundary conditions
will affect the physics in the center of the cylinder, even in the limit of an infinitely long one. 
Similar dimer coverings and RVB wavefunctions can be considered on cylinders with a kagome 
lattice (see e.g. Fig.~\ref{Fig:RVB_PlusMinus}). In that case, singlets are all oriented clockwise in both 
left and right triangles. It is known that RVB wavefunctions always exhibit short-range 
spin-spin correlations in two-dimensions (2D) although dimer-dimer correlations 
can be either short-range (kagome) or critical (square lattice) as mentioned above. 

\subsection{Topological sectors}

\begin{figure}
\includegraphics[width=1.0\columnwidth]{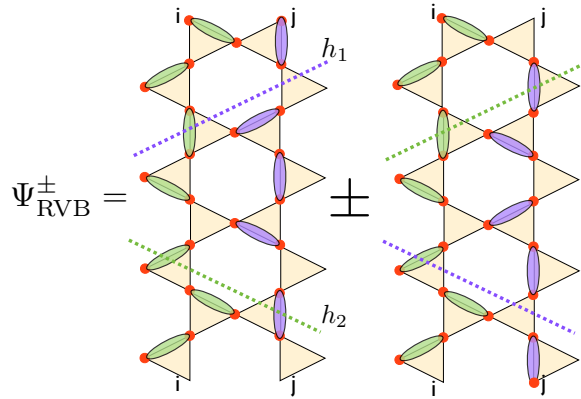}
\caption{Two valence bond configurations on a $4\times 2$ cylinder ($N_v=4$). The two configurations are obtained from each other by translating all 
dimers (in purple) along a (single) closed loop encircling the cylinder.
Such configurations can be distinguished from the parities $G_h=\pm 1$ of the number of dimers cut by {\it open} lines
along the $h_1$ and $h_2$ directions joining the two $B_L$ and $B_R$ ends of the cylinder and, hence, define two different topological sectors. 
Two RVB (variational) ground-states with $\big<G_h\big>=0$ can be constructed as equal-weight superpositions of all dimer coverings with $+$ or $-$ relative signs
between the two topological sectors. }
\label{Fig:RVB_PlusMinus}
\end{figure}

\begin{figure}
\includegraphics[width=1.0\columnwidth]{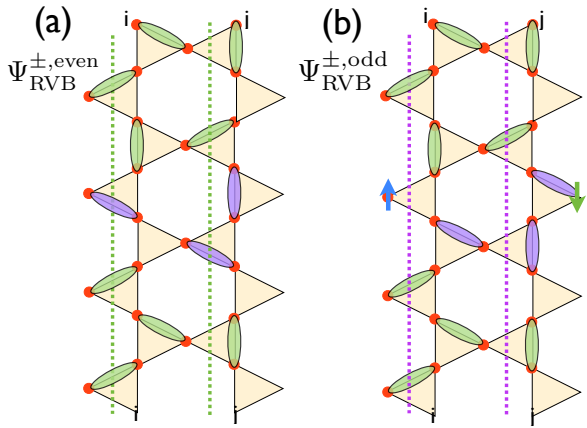}
\caption{
Two valence bond configurations on a $4\times 2$ cylinder ($N_v=4$ even). The two configurations are obtained from each other by translating all 
dimers (in purple) along a (single) {\it open} loop joining the two $B_L$ and $B_R$ ends of the cylinder (and adding extra spins).
Such configurations can be distinguished from the parity $G_v=\pm 1$ 
of the number of dimers cut by any {\it closed} loop
winding around the cylinder along the vertical direction and, hence, define two different ``even" and ``odd" topological sectors
(and the corresponding RVB states). 
}
\label{Fig:RVB_NvEVEN}
\end{figure}

\begin{figure}
\includegraphics[width=1.0\columnwidth]{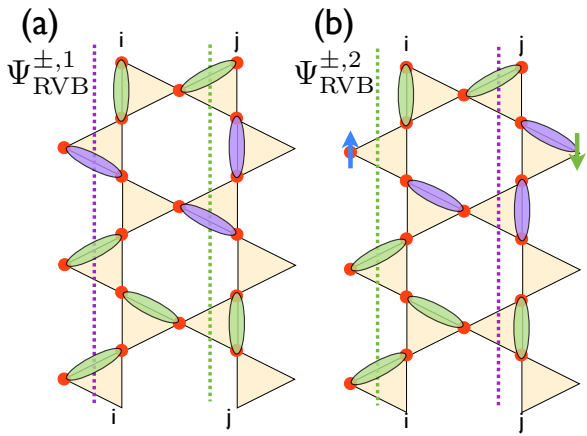}
\caption{Same as Fig.~\protect\ref{Fig:RVB_NvEVEN} for a $3\times 2$ cylinder ($N_v=3$ odd). 
}
\label{Fig:RVB_NvODD}
\end{figure}

Here we briefly review the crucial concept of topological sectors in the space of (hard-core)
dimer coverings (focusing on the kagome lattice)
 and show that four RVB wavefunctions belonging to 
different topological sectors can be constructed
on $N_v\times N_h$ cylinders with periodic (open and generalized) boundary conditions 
in the vertical (horizontal) direction when $N_v$ is even.  The case of odd perimeter will also be discussed. For illustration, small $4\times 2$ and $3\times 2$ 
cylinders are drawn for simplicity in Figs.~\ref{Fig:RVB_PlusMinus}, \ref{Fig:RVB_NvEVEN}, and \ref{Fig:RVB_NvODD} but our arguments are valid for any system size. 

Let us first consider the case of a cylinder with $N_v$ even.
Topological sectors can be defined by considering (i) a closed loop in the vertical direction 
winding around the cylinder (see Fig.~\ref{Fig:RVB_PlusMinus}) and 
(ii) two open lines along the crystal directions $h_1$ and $h_2$ at 30$^o$ angles w.r.t. the 
horizontal axis (see Fig.~\ref{Fig:RVB_NvEVEN}), joining the
two open ends $B_L$ and $B_R$ of the cylinder. As shown in Figs.~\ref{Fig:RVB_PlusMinus}, \ref{Fig:RVB_NvEVEN}, for a given configuration, the parities of the numbers of dimers cut
by these loops are conserved quantities under translation of the vertical loop (horizontal lines) 
along the horizontal direction
(vertical direction). Since the product of the three parities is constrained to be either even or odd
(depending on the choice of $N_h$ and $N_v$), NN dimer configurations
can be grouped into four disconnected sectors. Their are ``topological" in nature since 
any {\it local} Hamiltonian acting on the space of dimer configurations preserves the sectors. 

It is interesting to note that one can go from one topological 
sector to the other by {\it non-local} moves of dimers. 
For example, let us consider the left configuration of Fig.~\ref{Fig:RVB_NvEVEN}. By translating any staggered arrangement of dimers around a closed loop winding around  the cylinder 
by one lattice spacing, one permutes (changes) the parities $G_h$
measured along $h_1$ and $h_2$ for $N_h=4p+2$ ($N_h=4p$). Because the space of NN dimer 
coverings is divided into two disconnected sectors (fixing OBC), two RVB 
states can first be constructed separately in each sector. 
Such states should have the same energy density in the middle of the cylinder
(for a generic local su(2)-Hamiltonian) since nothing can distinguish the two states locally. 
However, on a {\it finite} cylinder, such RVB states don't have
the lowest variational energy since they break the mirror symmetry w.r.t. the horizontal direction (a symmetry assumed 
for the Hamiltonian) when $N_h=4p+2$.
However, by taking their superpositions both with relative plus or minus signs (see Fig.~\ref{Fig:RVB_PlusMinus}),
two appropriate variational GS $\Psi_{\rm RVB}^+$ and $\Psi_{\rm RVB}^-$ can be defined
(strictly orthogonal for $N_h=4p+2$).
Interestingly, starting from $\Psi_{\rm RVB}^+$, one can pictorially obtain $\Psi_{\rm RVB}^-$ by inserting a ``vison" line going all the way from the left to the right boundaries of the cylinder e.g. along the $h_1$ direction~: the vision operator counts the number of dimers cut by the line and adds a minus sign to the wave function for an odd number of cuts. 
In other words, the $\Psi_{\rm RVB}^+$ (no-vison) and $\Psi_{\rm RVB}^-$ (vison) are states with a 
{\it definite} $\mathbb{Z}_2$ flux through the cylinder. 

The two states $\Psi_{\rm RVB}^+$ and $\Psi_{\rm RVB}^-$ have been constructed for specific OBC for $B_L$ and $B_R$.
Shifting  by one lattice spacing a line of staggered dimers joining the two ends of the cylinder,
will change the parity $G_v$ of the numbers of dimers cut by loops winding around the cylinder, hence providing a change from, let say, the ``even" to the
``odd" topological sector, as seen in Fig.~\ref{Fig:RVB_NvEVEN}.
By applying this second type of non-local move to the two previous $\Psi_{\rm RVB}^+$ and $\Psi_{\rm RVB}^-$ wavefunctions, 
one can then construct 
four orthogonal variational RVB wavefunctions denominated as $\Psi_{\rm RVB}^{+,{\rm even}}$, $\Psi_{\rm RVB}^{-,{\rm odd}}$,
$\Psi_{\rm RVB}^{+,{\rm even}}$ and $\Psi_{\rm RVB}^{-,{\rm odd}}$.

Let us now briefly discuss the case of ``odd" cylinders i.e. cylinders with an odd number $N_v$ of unit cells. As shown in 
Fig.~\ref{Fig:RVB_NvODD}, the parity of the number of dimers cut by {\it closed} loops
encircling the cylinder along the vertical direction alternates along the cylinder.
This indicates that two consecutive columns become non-equivalent and the system spontaneously
dimerizes in the cylinder direction. By shifting a horizontal line of staggered dimers as before, one switches the parity of the 
``even" and ``odd" columns. This defines two disconnected classes of configurations from which
two related RVB states $\Psi_{\rm RVB}^{+,1}$ and $\Psi_{\rm RVB}^{+,2}$ can be constructed as equal weight superposition
of all dimer configurations of each class. 
In the center of long (enough) cylinders, these two RVB states are simply related by a unit translation along the cylinder.
Of course, as before, a vision line can be inserted between the two ends of the cylinder to derive two new 
$\Psi_{\rm RVB}^{-,1}$ and $\Psi_{\rm RVB}^{-,2}$ wavefunctions.

We finish this Section with the case of the square lattice. Because of the much more constrained nature of 
dimer configurations on the square lattice, one can construct an extensive number $\propto N_v$ of
topological sectors. This will be discussed in more details in Sec.~\ref{Sec:boundary}.

\section{PEPS representation of RVB states} 
\label{Sec:peps}

\subsection{Mathematical construction}

\begin{figure}
\begin{center}
 \includegraphics[width=0.8\columnwidth]{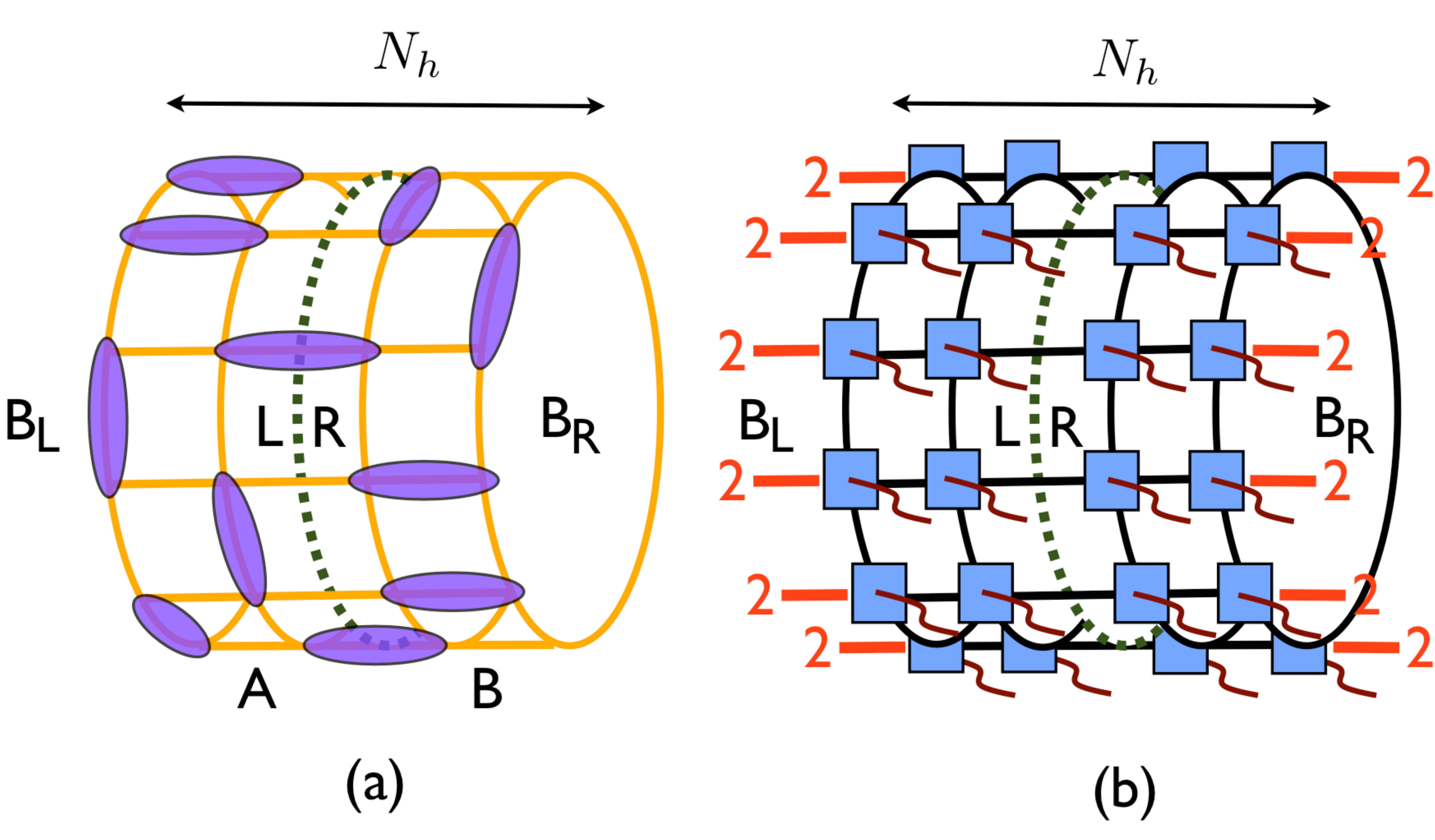}
 \includegraphics[width=0.8\columnwidth]{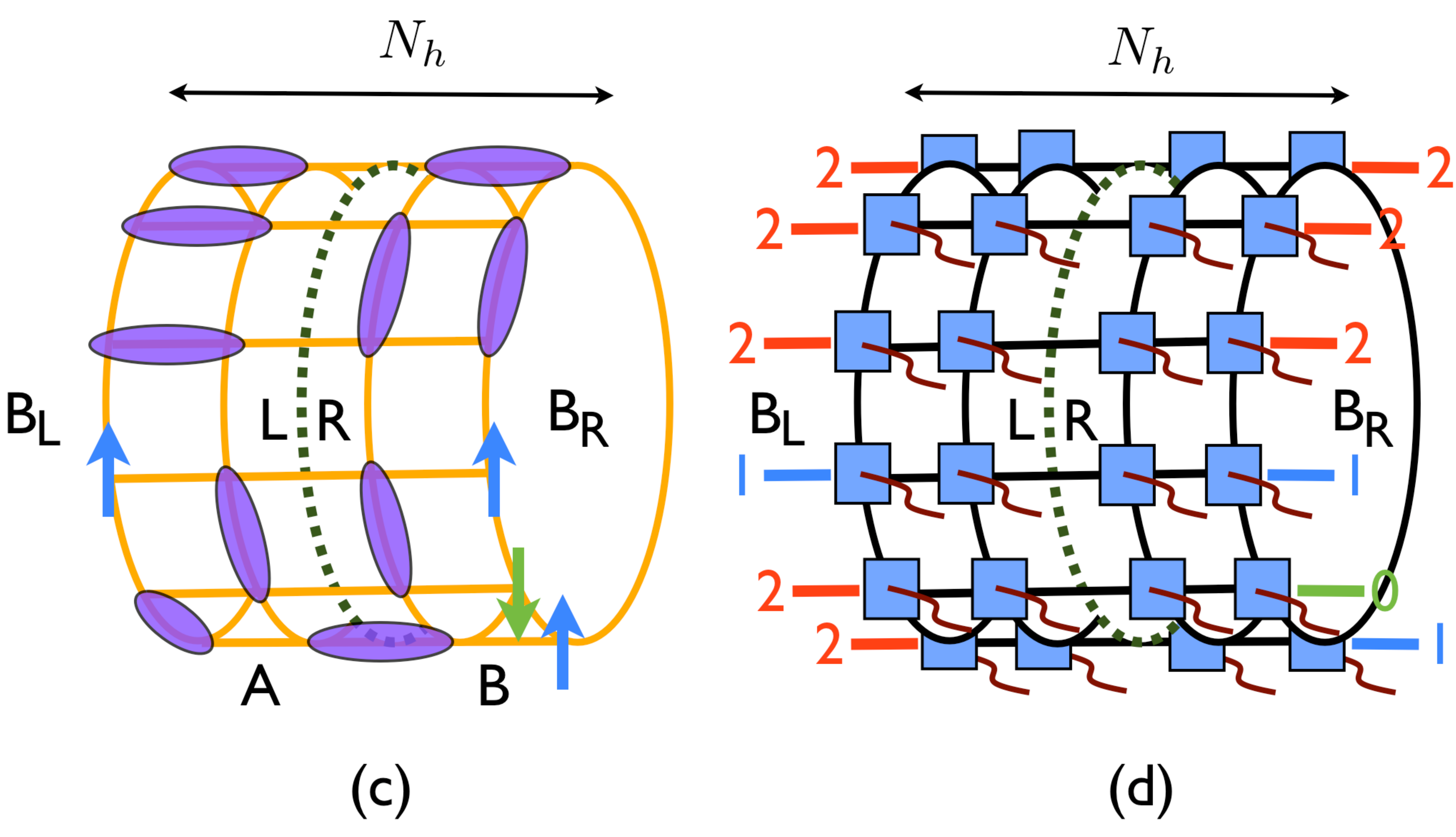}
 \end{center}
\caption{(Color online)
RVB wavefunctions on a cylindrical geometry:
equal-weight superposition of hardcore-dimer coverings [see e.g. (a,c)] 
have simple representations in terms of
PEPS (b,d). The $B_L$ and $B_R$ boundary conditions 
of Fig.~\ref{Fig:cylinders}(a,b) can be realized by fixing the virtual variables going out of the cylinder ends;
OBC (a) are defined  by setting all boundary indices to ``2" (b). 
Generalized boundary conditions (c) translate in the PEPS language by setting the boundary indices to 0 (spin $\downarrow$)
or 1 (spin $\uparrow$) (d). A bipartition of the cylinder
generates two L and R edges along the cut. }
\label{Fig:cylinders_peps}
\end{figure}

\begin{figure}\begin{center}
 \includegraphics[width=1.0\columnwidth]{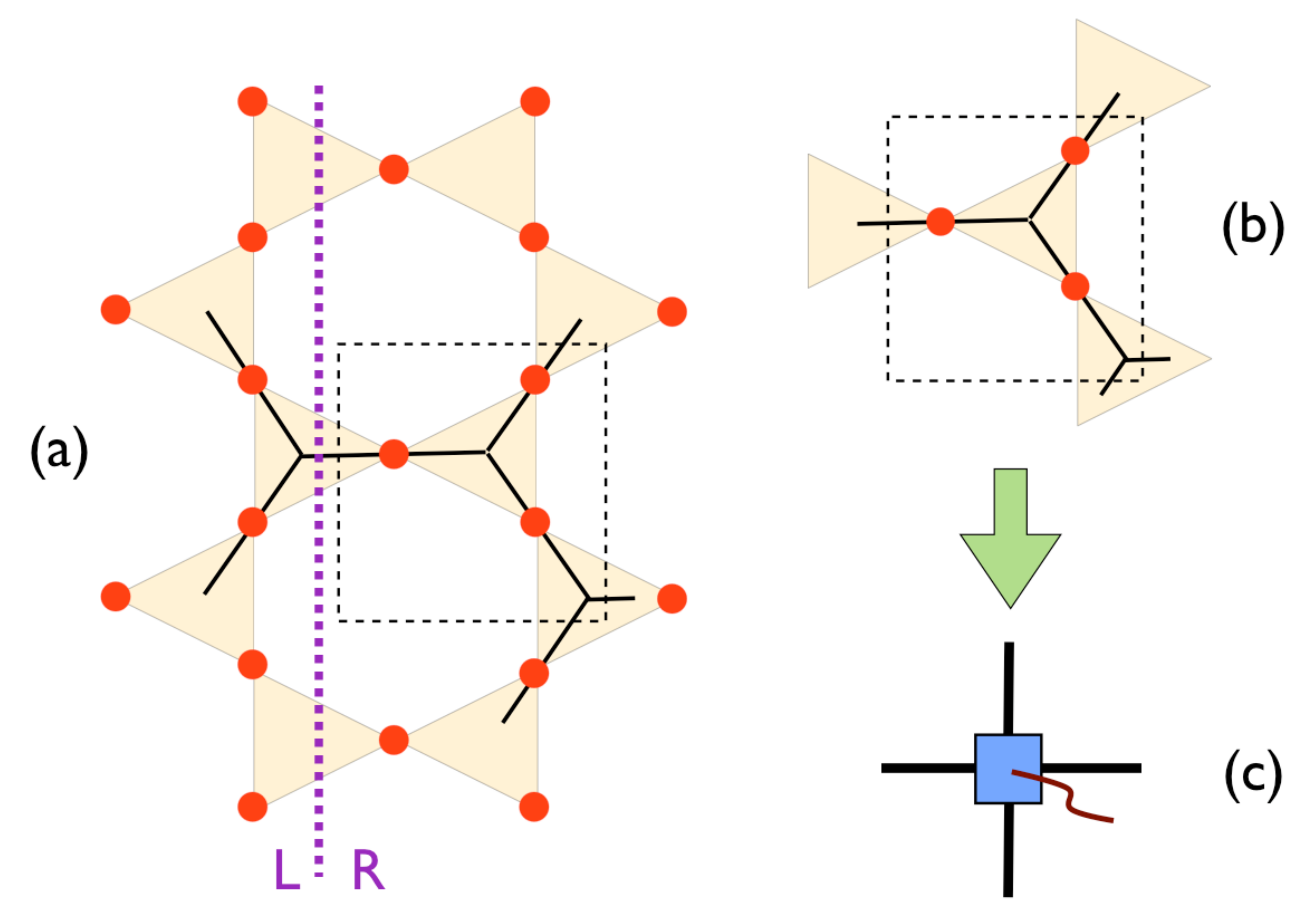}
 \end{center}
\caption{(Color online)
On the kagome lattice, an effective rank-5 tensor is constructed on each 3-site unit cell. 
Three site tensors (red dots) carrying the physical indices and two 120-degree tensors (in the center of the shaded triangles) are grouped together (a,b) to construct the basic tensor (c). The kagome lattice is then mapped onto an effective square lattice.
A partition of the cylinder in the vertical direction generates L and R edges (thick dotted line).}
\label{Fig:kagome}
\end{figure}

We start with the square lattice RVB wavefunction (NN  
\hbox{$|\uparrow\downarrow\rangle-|\downarrow\uparrow\rangle$} singlets are all oriented from one sublattice to the other) 
on a cylinder of length $N_h$ and circumference $N_v$, 
depicted in Fig.~\ref{Fig:cylinders_peps}(a,c), corresponding to an equal weight (and equal sign) summation of 
all (singlet) dimer coverings.
The RVB wavefunction can be expanded in the local $S_z$-basis,
$|\Psi_{\rm RVB}\rangle = \sum_{S} c_{S}|s_1,s_2,\ldots,s_M\rangle$, where $s_n=0,1$ are  
qubits (representing the two $S_z=\pm 1/2$ spin components) 
on the $M=N_h N_v$ sites and $S=\{s_n\}$. Such a state can in fact be
represented by a $D=3$ PEPS~\cite{verstraetewolf06,long_paper} (up to local unitaries) where
each lattice site is replaced by a rank-5 tensor $A^{s}_{\alpha,\alpha';\beta,\beta'}$
labeled by one physical index, $s=0$ or $1$,
and by four virtual bond indices (varying from 0 to 2) along the horizontal ($\alpha,\alpha'$) and vertical ($\beta,\beta'$) directions.
Physically, the absence of singlet on a bond is encoded by the virtual index being "2" on that bond. 
To enforce the hardcore dimer constraint, one takes
$A^{s}_{\alpha,\alpha';\beta,\beta'}=1$ whenever three virtual indices equal 2
and the fourth one equals $s$, and $A^{s}_{\alpha,\alpha';\beta,\beta'}=0$ otherwise. 
The amplitudes $c_S$ are then
obtained by contracting all virtual indices, except the ones at the ends of the cylinder 
fixed by boundary conditions, as depicted in Fig.~\ref{Fig:cylinders_peps}(b,d). 
For the kagome lattice, as shown in Fig.~\ref{Fig:kagome}, the RVB state can be represented in terms of 
rank-3 tensors, (i) $A^{s}_{\alpha;\beta}$ on the sites -- $A^{s}_{2;s}=A^{s}_{s;2}=1$ and zero otherwise -- and 
(ii) on the center of each triangle, $R_{2,2,2}=1$, 
and $R_{\alpha,\beta,\gamma}=\epsilon_{\alpha\beta\gamma}$ otherwise, 
with $\epsilon_{\alpha\beta\gamma}$ the antisymmetric tensor.~\cite{long_paper} 
One can then group the 3 sites 
on each unit cell to obtain a rank-5 tensor (the physical dimension is now $2^3=8$) connected on an effective square lattice (Fig.~\ref{Fig:kagome}(b,c)). 
Note that for the Kagome PEPS, one can find a local parent Hamiltonian for 
which the degeneracy is equal to 4 on the torus.~\cite{long_paper}

In the PEPS formulation the boundary conditions $B_L$ and $B_R$ can be simply set by fixing the 
virtual states on the bonds ``sticking out" at each cylinder end.
E.g. open boundary conditions as in Fig.~\ref{Fig:cylinders_peps}(a) are obtained by setting 
the boundary virtual indices to ``2" as shown
in Fig.~\ref{Fig:cylinders_peps}(b). Generalized boundary conditions can be realized as in Fig.~\ref{Fig:cylinders_peps}(c,d) by setting some of the virtual indices on the ends to 0 or 1.

\subsection{Topological energy splittings of kagome RVB wavefunctions}

\begin{figure}
\includegraphics[width=0.9\columnwidth]{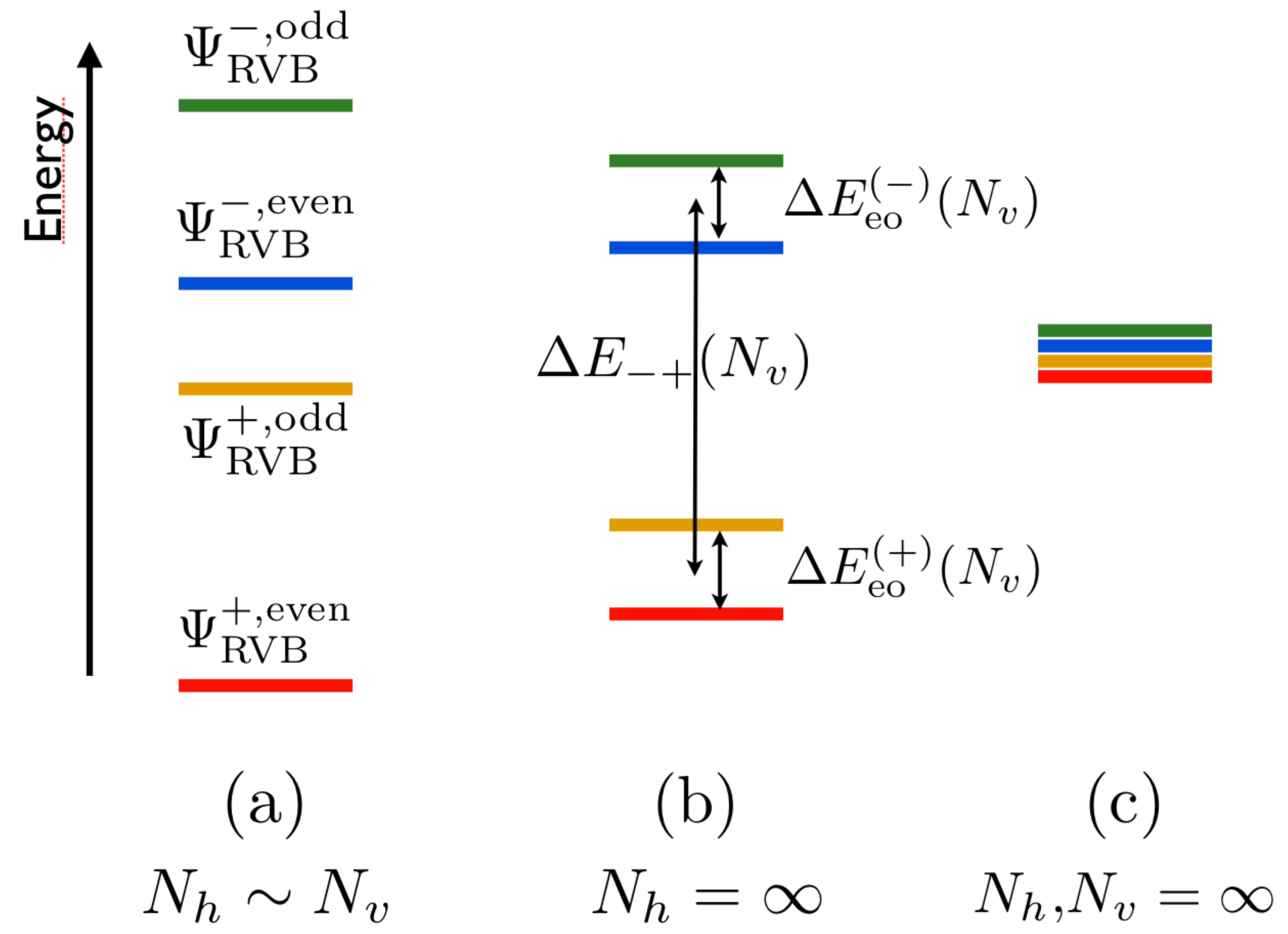}
\caption{Illustration of the energy splitting between the four (variational)
RVB wavefunctions for the kagome HAF. From left to right, the cylinder length (at fixed perimeter $N_v$ even)
and then, its perimeter are increased to infinity. 
}
\label{Fig:topo-gap}
\end{figure}

More and more numerical data from DMRG simulations support
the claim that the NN quantum HAF on the kagome lattice is a 
topological $\mathbb{Z}_2$ 
spin liquid~\cite{White,Balents2,Schollwoeck}. It is therefore interesting (and relevant) to consider the previous topological NN-RVB wavefunctions
as variational ground-state ans\"atze for the NN HAF Hamiltonian on the kagome lattice,
\begin{equation}
H=J\sum_{\big< ij\big>} {\bf S}_i\cdot {\bf S}_j\, ,
\end{equation}
where $\bf S_i$ is the spin-1/2 operator at site $i$ and $\big< ij\big>$ stands for all NN bonds of the
kagome lattice and the exchange constant is been set to $J=1$ from now on.
Although (i) the (local) parent Hamiltonian of the NN-RVB wavefunction contains much more complicated 
interactions~\cite{long_paper} and, reversely, (ii) the ground-state of the NN HAF is far more involved 
that a simple NN-RVB (e.g. containing singlets bonds beyond NN),
we believe generic features on the finite size energy splitting between the different topological sectors
(topological gap) can be obtained by using simple NN RVB wave functions. 
 A schematic picture in Fig.~\ref{Fig:topo-gap} illustrates the expected GS multiplet structure
for increasing system size. In the 2D thermodynamic limit, when both cylinder length and perimeter are infinite,
one expects all energy splittings to vanish and the GS to become four-fold degenerate.  

\begin{figure}
\includegraphics[width=1.0\columnwidth]{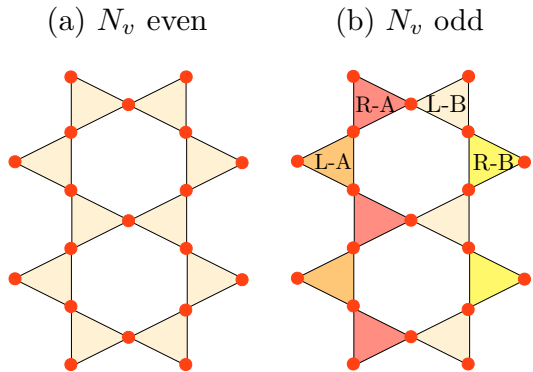}
\caption{Schematic patterns of the exchange interaction on the triangles in the center of (quasi-) infinite cylinders.
For $N_v$ even ($N_v$ odd), the system is uniform (dimerized). 
}
\label{Fig:RVB_VBC}
\end{figure}

\begin{figure}
\includegraphics[width=1.0\columnwidth]{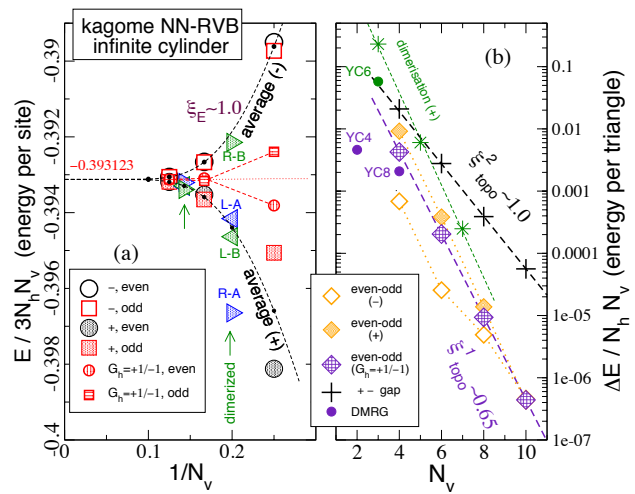}
\caption{(a) Finite size scaling of the energy (per site) of the four RVB wave functions on infinite kagome cylinders versus inverse perimeter $N_v$.
``even" and ``odd" refer to the parity of the number of spins frozen on the cylinder boundaries. ``$+$" and ``$-$" states differ by the absence or presence
of a vison horizontal line, respectively. The energies of the fixed parity ($G_h=\pm 1$) states are obtained by averaging the $+$ and $-$ energies (since the cross-terms vanish) separately in the even and odd sectors.
The energies of the four non-equivalent triangles of odd-perimeter infinite cylinders (with no vison) are also included. 
Averages over the even and odd energies separately in the no-vison ($+$) and vison ($-$) sectors are also shown.
(b) Corresponding energy splittings (normalized per 3-site unit cell) vs $N_v$ (see Fig.~\ref{Fig:topo-gap}(b)). 
We also include the dimerization energy of the $\Psi_{\rm RVB}^{+,1,2}$ states on odd-perimeter infinite cylinders,
defined as the energy difference between even and odd columns. DMRG data (S.~R.~White) for the dimerization of the YC6 cluster ($N_v=3$) 
or the spittings of the YC4 ($N_v=2$) and YC8 ($N_v=4$) clusters are shown for comparison. 
Dashed straight lines correspond to exponential fits of the form $A_0\exp{(-N_v/\xi)}$.
}
\label{Fig:EvsNv}
\end{figure}

The PEPS formalism allows to compute exactly the variational energy of the NN RVB wavefunctions on cylinders of perimeter  $N_v$ up to
$N_v=10$ and length $N_h\rightarrow\infty$. $\Psi_{\rm RVB}^{+,{\rm even}}$ is obtained using the local rank-5 tensors described 
above and OBC. To get $\Psi_{\rm RVB}^{-,{\rm even}}$ one inserts a
``vison" line joining the two boundaries of the cylinder by 
putting a string of $Z=\mathrm{diag}(1,1,-1)$ operators on the
bonds.~\cite{schuch:peps-sym} Finally,
$\Psi_{\rm RVB}^{\pm,{\rm odd}}$ are obtained by using GBC for $B_L$ and
$B_R$.  The energy is computed at the center of the cylinder after full
convergence with increasing cylinder length $N_h$ is reached (typically
$N_h\sim 10 N_v$ is enough). We have checked numerically that all states
possess mirror symmetry of the energy density w.r.t. the horizontal axis
(as expected from their symmetry) and are uniform (staggered) for $N_v$
even (odd) as illustrated in Fig.~\ref{Fig:RVB_VBC}.  Interestingly, for
$N_v=8$ and $N_v=10$ the lattice $C_{6v}$ symmetry around an hexagon center is almost
fully recovered (i.e. the
vertical and $30$-degree bonds become equivalent). The energy (per site) of the four orthogonal RVB wave functions
are plotted in Fig.~\ref{Fig:EvsNv}(a) vs $1/N_v$. After averaging the energies of the even and odd states, one obtains very accurate fits of the
exponentially fast convergence of the energies of the RVB wave functions in the ``$+$" and ``$-$" topological sectors, with a very
short characteristic length-scale $\xi_E\sim 1.0$. 
The extrapolated energy agrees very well with a recent estimate based on a Gutzwiller-projected superconducting
wavefunction~\cite{Yao_2012}. 
Although this variational energy is much higher than most recent
variational estimates~\cite{White,Balents2,Schollwoeck} (between $-0.437$ and $-0.439$), we believe the observed finite size behaviors and energy splittings (topological gaps)
should be generic of $\mathbb{Z}_2$ spin liquids. 
For example, we find that the average over the variational energies of
the four RVB 
wavefunctions exhibits surprisingly small size dependance, in striking 
correspondence with DMRG results~\cite{White,Schollwoeck}.

The splittings between the variational RVB GS  defined in Fig.~\ref{Fig:topo-gap}(b) are plotted using a logarithmic
scale in Fig.~\ref{Fig:EvsNv}(b)
as a function of the perimeter $N_v$ of the infinite cylinder. Exponential decay of the topological splittings vs $N_v$
are seen revealing {\it two} typical length-scales $\xi^1_{\rm topo}\sim 0.65$ and $\xi^2_{\rm topo}\sim 1.01$, associated to the
even-odd and $+-$ gaps, respectively. Note that the dimerization energy of the $\Psi_{\rm RVB}^{+,1}$ (or $\Psi_{\rm RVB}^{+,2}$)
states follows the same exponential decay as the even-odd topological gaps. For a very long cylinder with fixed boundary conditions, 
we therefore predict the following finite size scaling of the largest topological splitting (cost of inserting a horizontal vison line),
\begin{equation}
\Delta E_{+-}\simeq 1.06\, N_v N_h \exp{(-0.99\, N_v)}  \, .
\label{Eq:scaling1}
\end{equation} 
The cost of freezing an {\it odd} number of spins at the boundary is given by,
\begin{equation}
\Delta E_{eo}\simeq 1.95\, N_v N_h \exp{(-1.54\, N_v)}\, ,
\label{Eq:scaling2}
\end{equation} 
for the case where the state has a definite parity ($G_h=\pm 1$).
For a definite $\mathbb{Z}_2$ flux in the cylinder (e.g. a state with or without a vison), moderate corrections
occur for perimeter $N_v\le 8$ as seen in Fig.~\ref{Fig:EvsNv}(b).
In DMRG, the two different even and odd sectors can be fixed~\cite{White} by moving a site from one end of the cylinder to the
other, which would be the same as pinning sites with strong fields on either end. However, it is not clear whether 
the DMRG algorithm chooses a definite $G_h$ parity or a definite $\mathbb{Z}_2$ flux or none of the two.

\section{Boundary Hamiltonian on infinite cylinders}
\label{Sec:boundary}

\subsection{Bipartition and reduced density matrix} 

To define the boundary Hamiltonian of the RVB wavefunctions,
we partition the $N_v\times N_h$ cylinder into two half-cylinders
of lengths $N_h/2$,  as depicted in Fig.~\ref{Fig:cylinders_peps}. 
Partitioning the cylinder into two half-cylinders
(playing the role of two A and B subsystems as defined in the Introduction)
reveals two edges L and R along the cut. Ultimately, we
aim to take the limit of infinite cylinders, i.e. $N_h\rightarrow\infty$ as before. 

For a topological state, the boundary Hamiltonian 
depends on the choice of the wavefunction within the (variational) GS degenerate manifold.
In other words, it depends upon (i) the choice of the
$B_L$ and $B_R$ cylinder boundaries that impose the parity $G_v$  ($B_L$ and $B_R$ have to ``match") 
and (ii) the possible insertion of a horizontal vison line (or equivalently a
$\mathbb{Z}_2$ flux through the cylinder).  
For simplicity, we restrict ourselves to the $+$ combination of Fig.~\ref{Fig:RVB_PlusMinus} (no vison)
but still consider arbitrary choices of the boundary conditions at the ends of the cylinder.

\begin{figure}
\begin{center}
 \includegraphics[width=0.6\columnwidth]{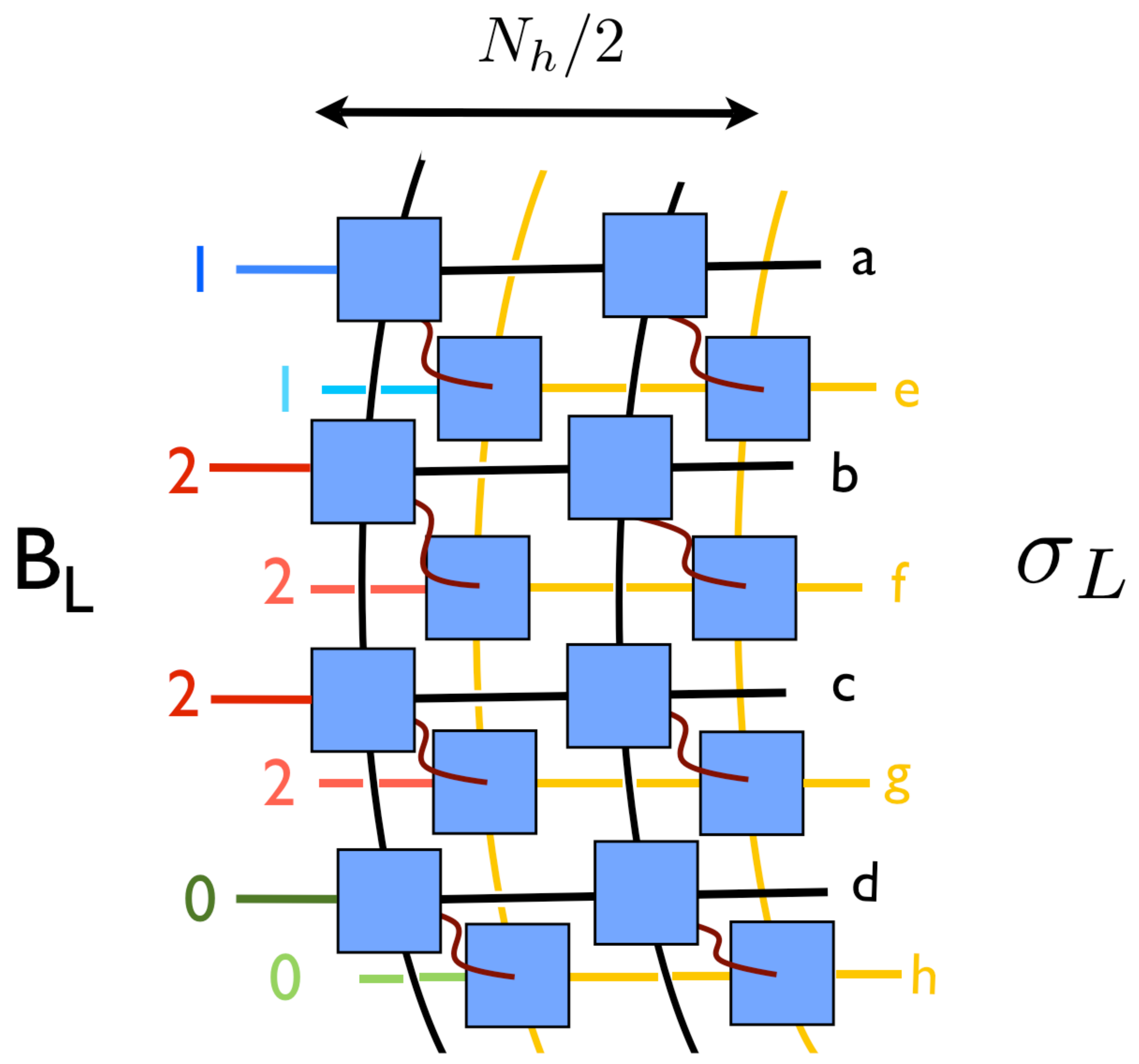}
 \end{center}
\caption{(Color online)
Boundary operator $\sigma_L$ obtained by contracting all physical indices (wavy lines
connecting the tensors in the front and in the back) of the left
half-cylinder. Here, arbitrary boundary conditions have been chosen for $B_L$.
}
\label{Fig:sigma}
\end{figure}

The boundary Hamiltonian ${\tilde H}_b$ can be derived from 
the reduced density operator 
$\sigma_b^2 = \exp{(- \tilde H_b)}$ acting on the edge indices, following the 
procedure given in Ref.~\onlinecite{ciracpoilblanc2011}. For the kagome lattice, there is no reflection 
symmetry w.r.t. the cut so
the RDM for the left (right) side takes the form 
$\sigma_{bL}^2=\sqrt{\sigma_{\! R}^{\, t}}\sigma_L\sqrt{\sigma_{\! R}^{\, t}}$ 
($\sigma_{bR}^2=\sqrt{\sigma_{\! L}^{\, t}}\sigma_R\sqrt{\sigma_{\! L}^{\, t}}$) where  
$\sigma_L$ and $\sigma_R$ are obtained by contracting the tensors of the left and right 
half-cylinders, respectively, as shown in Fig.~\ref{Fig:sigma} (see Ref.~\onlinecite{ciracpoilblanc2011}
for details).
Note that 
$\sigma_{bL}^2$ and $\sigma_{bR}^2$ give identical ES. For clarity, we restrict ourself to $\sigma_{bL}^2$.
Ultimately, we are interested in RVB cylinders 
with infinite lengths in both directions. First, we fix the cylinder
perimeter ($N_v=4,6,8$) and take the limit 
$N_h\rightarrow\infty$ as shown in Appendix A (in practice, the RDM for $N_h\sim 10N_v$ is fully converged).
The behaviors of the boundary Hamiltonian and the ES as a function of cylinder perimeter is then analyzed
(see Appendix B for explicit finite size scalings).

\subsection{Boundary Hamiltonian}

\subsubsection{Disconnected topological sectors in the PEPS representation}

The concept of boundary Hamiltonians is described in details in Ref.~\onlinecite{ciracpoilblanc2011} and, for topological states, in Ref.~\onlinecite{schuch2012}.
Here, we provide the details of their numerical computation for the RVB states.
A crucial feature of the topological states is that
the RDM depends intrinsically on the choice of the boundary conditions ($B_L=B_R$ for simplicity), {\it even when $N_h\rightarrow\infty$}.
Indeed, the configurations of virtual indices (on the horizontal bonds) of any
vertical column is split in two disjoined sectors which, due to local constraints, are conserved from column to column
(and hence can be addressed independently from proper choices of $B_L=B_R$). 
This is directly connected to the partition of the space of NN dimer coverings 
into disconnected (even and odd for the kagome lattice) topological sectors discussed earlier. 
Indeed, the number of $|2\big>$ states on a column of virtual bonds corresponds to the number of bonds with no dimers.
Therefore, for the kagome lattice, in a given topological sector, the parity of the number of virtual $|2\big>$
states on the columns is conserved from column to column. 
Consequently, the boundary Hamiltonian of the Kagome RVB wave function conserves the parity of the
number of $|2\big>$ states, as in Kitaev's toric code.  
On the square lattice, extra constraints impose the conservation of the difference between the number
of $|2\big>$ states between two alternating sublattices on the edge.
Hence, although the RDM (and $\tilde H_b$) acts 
on all $3^{N_v}$ degrees of freedom of the $L$ (or $R$) edge, 
in each sector it contains a
finite fraction of zero-weight eigenvalues, i.e. a finite
fraction of eigenstates of $\tilde H_b$ have infinite-energy.
Calling  ${\cal P}$ the projector on the finite-energy subspace,
we split $\tilde H_b$ as 
$\tilde H_b=H_1 + \beta_{\rm topo} ({\rm Id} - {\cal P})$,
where Id is 
the $3^{N_v}\times 3^{N_v}$ identity-operator, $\beta_{\rm topo}\rightarrow\infty$,
and $H_1$ is supported by the non-zero
eigenvalues sector of the RDM. 
More precisely, $H_1$ can be factorized as $H_1=H_{\rm local}{\cal P}$
where $H_{\rm local}$ is a Hamiltonian (shown later to be local) acting on the 
whole boundary space~\cite{note2}, commuting with $\cal P$ and 
{\it independent  on BC}.
The Kagome cylinder has only two sectors defined by $\cal P=P_{\rm even}$
and ${\cal P}={\cal P}_{\rm odd}={\rm Id}-P_{\rm even}$, which can be obtained by 
choosing an even or odd number of ``2" external (virtual) indices for $B_L=B_R$, respectively.
On the square lattice cylinder, there are $N_v+1$ disconnected sectors defined
by projectors ${\cal P}_\Delta$ enforcing a fixed difference $N_{2,X}-N_{2,Y}=\Delta$ 
of the numbers of ``2" on two X and Y alternating sublattices on the edge, 
$\Delta=-N_v/2,\cdots,N_v/2$. 
The fact that $H_1$ is known for all
sectors implies that $H_{\rm local}$ is uniquely determined
as will be shown in the next Subsection.

\subsubsection{Practical derivation of $H_{\rm local}$}

In practice, each numerical calculation is done for a specific choice of the boundary conditions $B_L$ and $B_R$
(for simplicity, we assume here $B_L=B_R$) on the cylinder ends which
determines a given conserved sector (mathematically characterized by some projector ${\cal P}$),
support of the corresponding boundary Hamiltonian.
Reversely, all sectors (associated to different projectors $\cal P$) 
can be obtained from proper choices of the boundary conditions $B_L=B_R$
(like the sectors defined by the projectors
${\cal P}_{\rm even}$ and ${\cal P}_0$ which can be addressed by choosing OBC). 

On the Kagome cylinder, the two sectors defined by $\cal P=P_{\rm even}$
and ${\cal P}={\cal P}_{\rm odd}={\bf 1}^{\otimes N_v}-P_{\rm even}$ can be obtained by 
choosing even (e.g. OBC) or odd number of ``2" external (virtual) indices on both the left and right boundaries 
of the cylinder, respectively. 
We can then construct a ``mixed" RDM (for the right part),
\begin{equation}
\sigma_{b}^2=\sqrt{\sigma_{\! L}^{\, t}}\sigma_R\sqrt{\sigma_{\! L}^{\, t}},
\label{Eq:sigma_mixed}
\end{equation}
by considering the linear superpositions
\begin{eqnarray}
\sigma_R&=&\sigma_{R,{\rm even}}+\sigma_{R,{\rm odd}},\nonumber\\
\sigma_L&=&\sigma_{L,{\rm even}}+\sigma_{L,{\rm odd}}, \label{Eq:sum_sigma}
\end{eqnarray}
where $\sigma_{T,{\rm p}}$ are obtained by contracting the left ($T=L$) and right ($T=R$) half-cylinders 
(see Ref.~\onlinecite{ciracpoilblanc2011}) with appropriate p=``even" or p=``odd" parity boundary conditions
and the ``equal weight" normalization condition, 
\begin{equation}
{\rm Tr}\{(\sigma_{L,{\rm even}})^t\sigma_{R,{\rm even}}\}
={\rm Tr}\{(\sigma_{L,{\rm odd}})^t\sigma_{R,{\rm odd}}\}=1.
\end{equation}
Since $\sigma_{T,{\rm even}}$ and $\sigma_{T,{\rm odd}}$ are supported on disconnected subspaces, 
the mixed RDM splits into orthogonal contributions,
$
\sigma_b^{\, 2}=\rho_{\rm even}+\rho_{\rm odd},
$
where
\begin{eqnarray}
\rho_{\rm even}&=&\sqrt{\sigma_{\! L,{\rm even}}^{\,\, t}}\, \sigma_{R,{\rm even}}\, \sqrt{\sigma_{\! L,{\rm even}}^{\,\,t}}\, ,\nonumber\\
\rho_{\rm odd}&=&\sqrt{\sigma_{\! L,{\rm odd}}^{\,\, t}}\, \sigma_{R,{\rm odd}}\, \sqrt{\sigma_{\! L,{\rm odd}}^{\,\,t}}\, .
\label{Eq:rho_eo}
\end{eqnarray}
Since $\sigma_b^{\, 2}$ is supported by the whole Hilbert space, $H_{\rm local}$ can be {\it uniquely} defined by 
setting $\sigma_b^{\, 2}=\exp{(-H_{\rm local})}$,
enabling a direct computation of $H_{\rm local}= -\ln{\sigma_{b}^2}$ from Eqs.~(\ref{Eq:sigma_mixed}) and (\ref{Eq:sum_sigma}). 
Reversely, the generic form of the boundary Hamiltonian associated to the (normalized)
RDM $\rho_p$ is given by,
\begin{equation}
\tilde H_b=H_{\rm local}{\cal P}_p+\beta_{\rm\infty} {\cal P}_{\bar p}\, ,
\end{equation}
with $\beta_{\rm\infty}\rightarrow\infty$ and where $p$ ($\bar p$) refers to the "even" ("odd") or "odd" 
("even") parity sector.

\begin{figure}\begin{center}
 \includegraphics[width=1.0\columnwidth]{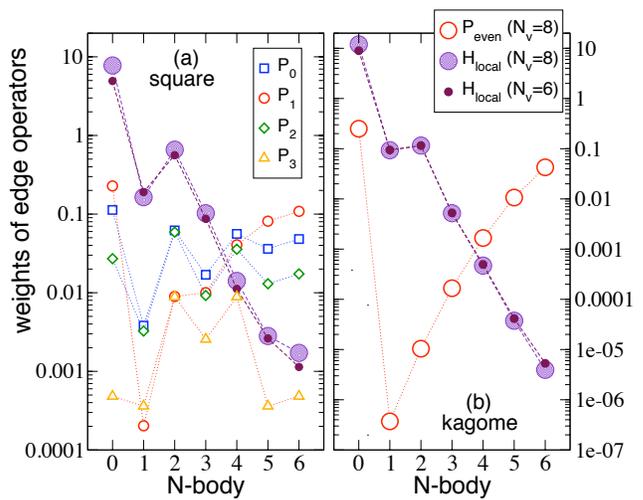}
 \end{center}
\caption{(Color online)
Weights of the projectors ${\cal P}_\Delta$, $\Delta=0,1,2$ and $3$ (square lattice, $N_v=6$) 
and ${\cal P}_{\rm even}$ (kagome lattice, $N_v=8$)
expended in terms of N-body operators. Same for $H_{\rm local}$ for the RVB wavefunction on the
square (a) and kagome (b) lattices, on $N_v=6$ (small symbols) and $N_v=8$ (large symbols)
infinite cylinders.
}
\label{Fig:weights_2}
\end{figure}

For the RVB wavefunction on the square lattice, there are $N_v/2+1$ orthogonal sectors defined
by the projectors ${\cal P}_\Delta$ enforcing a fixed difference $|N_{2,A}-N_{2,B}|=\Delta$ 
on the two A and B alternating sublattices on the one-dimensional edge, 
$\Delta=0,1,\cdots,N_v/2$. We show (for $N_v=6$) the expansion of these projectors in terms of N-body operators in Fig.~\ref{Fig:weights_2}(a)), highlighting clearly their highly non-local character. 
We construct the ``mixed" RDM 
$\sigma_{b}^2$ from the linear superposition
\begin{equation}
\sigma_b=\sum_{\Delta=0}^{N_v/2}\sigma_{\Delta}\, ,
\label{Eq:sum_sigma2}
\end{equation}
where $\sigma_{\Delta}$ is obtained by contracting the infinite
(left or right) half-cylinder 
(see Ref.~\onlinecite{ciracpoilblanc2011}) with appropriate boundary conditions
and normalized according to ${\rm Tr}(\sigma_{\Delta}^2)=1$. 
The operators $\sigma_{\Delta}$ are supported on orthogonal subspaces which span the whole
space of virtual indices on the edge i.e. $\sum_\Delta {\cal P}_\Delta={\bf 1}^{\otimes N_v}$.
Therefore, since $\sigma_{b}^2$ lives on the whole Hilbert space of the edge, one can {\it uniquely}
define $H_{\rm local}$ as $H_{\rm local}=-\ln{\sigma_{b}^2}$. Consequently, $H_{\rm local}$ can be computed 
numerically using Eq.~\ref{Eq:sum_sigma2}.
It also follows that, for each sector, the corresponding boundary Hamiltonian is,
\begin{equation}
\tilde H_b=H_{\rm local}{\cal P}_\Delta+\beta_{\rm \infty} {\bar{\cal P}}_{\Delta}\, ,
\end{equation}
with $\beta_{\rm \infty}\rightarrow\infty$ and ${\bar{\cal P}}_{\Delta}={\bf 1}^{\otimes N_v}-{\cal P}_\Delta$ 
is the projector on the complementary subspace.

\subsubsection{Expansion in terms of N-body operators: numerical results}

Next, we wish to explore the non-local/local characters of the $H_1$/$H_{\rm local}$ edge operators.
Any operator ${\cal O}_{\rm edge}$ acting on the edge can be expanded in terms of $3^{2 N_v}$
orthogonal operators. For this purpose, 
we use a local basis of 9 (normalized)
operators $\{ {\hat x}_0,\cdots, {\hat x}_8\}$ which act 
on the local basis of configuration (at some site $i$), $\{ |0\big>, |1\big>, |2\big> \}$, e.g.
${\hat x}_0={\bold 1}$, ${\hat x}_1=\sqrt{\frac{3}{2}}(|0\big>\big<0| -|1\big>\big<1|)$ 
and ${\hat x}_2=\frac{1}{\sqrt{2}}(|0\big>\big<0|+|1\big>\big<1|-2|2\big>\big<2|)$,
for the diagonal matrices, 
complemented by  
$\hat x_3=\hat x_4^\dagger=\sqrt{3}|0\big>\big<1|$
acting as ``spin" operators,
and $\hat x_5=\hat x_7^\dagger=\sqrt{3} |2\big>\big<0|$ and 
$\hat x_6=\hat x_8^\dagger=\sqrt{3} |2\big>\big<1|$
acting as annihilation and creation (hardcore) bosonic operators.~\cite{note0}
The expansion in terms of N-body operators reads (see Appendix C for more details),
\begin{eqnarray}
{\cal O}_{\rm edge} &=& c_0 N_v+\sum_{\lambda,i} c_{\lambda} {\hat x}_\lambda^i 
+ \sum_{\lambda,\mu,r,i} d_{\lambda\mu}(r) \, {\hat x}_\lambda^i {\hat x}_\mu^{i+r} \nonumber \\
&+& \sum_{\lambda,\mu,\nu,r,r',i} e_{\lambda\mu\nu}(r,r') \, {\hat x}_\lambda^i {\hat x}_\mu^{i+r} {\hat x}_\nu^{i+r'} 
+ \cdots \, ,\label{Eq:Hb0}
\end{eqnarray}
where each group of terms involves products of $N$ ($1\le N\le N_v$) on-site ${\hat x}_\lambda$ 
($\lambda\ne 0$) operators.
Here the sums are restricted to non-equivalent
relative distances and only translations giving {\it distinct} sets of sites are performed. 
The (real) coefficients appearing in (\ref{Eq:Hb0}) have been computed 
for ${\cal O}_{\rm edge}={\cal P}_{\rm even}$ (Kagome lattice), ${\cal O}_{\rm edge}={\cal P}_\Delta$ (square lattice)
and ${\cal O}_{\rm edge}=H_{\rm local}$ (Kagome and square lattices)
on infinitely-long cylinders of perimeters $N_v=6$ and $N_v=8$ up to order $N=6$.

As seen from the distribution of their weights in Figs.~\ref{Fig:weights_2}(a,b), projectors 
are highly non-local, conferring a fundamentally non-local character to the boundary Hamiltonian $H_1$.
This is also to be expected for {\it realistic} topological GS of microscopic 
Hamiltonians on geometries involving open or fixed BC in some directions

The total weight corresponding to each order of the expansion of $H_{\rm
local}$ in terms of N-body operators are shown in
Fig.~\ref{Fig:weights_2}(a,b) as a function of the order $N$.  Finite size
effects are remarkably small and we believe the results for $N_v=8$ are
converged.  The data reveal clearly an exponential decay of the weight
with the order $N$.
In other words, $H_{\rm local}$ contains primarily one- and two-body contributions
(in addition to the the normalization constant). This is the first part of the proof that 
$H_{\rm local}$ is indeed local. However, one still needs to go beyond the analysis
and investigate further the r-dependence of the leading two-body contributions.
In the next Subsection, we show that $H_{\rm local}$  
of the $\mathbb{Z}_2$ topological RVB is basically a short-range two-body Hamiltonian.  
In contrast, the RVB wave function on the square lattice exhibits
a long-range two-body potential term.

\begin{figure}\begin{center}
 \includegraphics[width=1.0\columnwidth]{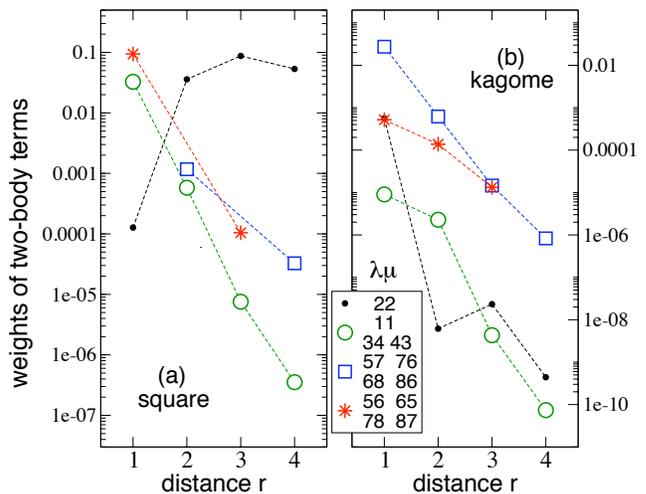}
 \end{center}
\caption{(Color online)
Weights $|d_{\lambda\mu}(r)|^2$ of the  2-body operators appearing in $H_{\rm local}$
for the square (a) and kagome (b) lattices. 
In (a) the diagonal interaction $\hat x_2^i \hat x_2^{i+r}$ shows a long-range behavior.
}
\label{Fig:two-body}
\end{figure}

\subsubsection{Local boundary Hamiltonian: an effective one-dimensional t--J model}

Next, we investigate the exact connection between the boundary Hamiltonian and the bulk 
properties of the system. 
We look for its explicit form, trying to make the connection with $D=3$ models, 
with su(2) symmetry corresponding to the $1/2\oplus 0$ representation. 

The boundary Hamiltonian belongs to the $1/2\oplus 0$ representation of su(2) and
its Hilbert space is the same as the one of a  {\it bosonic} t--J model.
Therefore, $H_{\rm local}$ is formally equivalent to a 
one-dimensional (1D) ``t--J model"~\cite{tJ}
describing motion of (bosonic) ``holes" (the ``2") in a spin fluctuating background 
(the ``0" and ``1" qbits) supplemented by additional density-density and
pair-field terms, conferring a superfluid character to the edge.
One can then rewrite 
the previous local operators in the notations of the 1D 
t--J model~\cite{tJ}.
We define bosonic creator operators $b_{i,s}^\dagger$ ($s=0,1$)
of the $|0\big>$ and $|1\big>$ states 
from the ``vacuum" $|2\big>$ (at some site $i$)
as $b_{i,s}^\dagger = |s\big>\big< 2|$, which naturally enforce the local Gutzwiller constraint of no doubly-occupied
site (in terms of hardcore bosons), so we can identify e.g. $b_{i,0} = \frac{1}{\sqrt{3}} {\hat x}_5$, $b_{i,1} = \frac{1}{\sqrt{3}}{\hat x}_6$, 
$b_{i,0}^\dagger = \frac{1}{\sqrt{3}}{\hat x}_7$ and $b_{i,1}^\dagger = \frac{1}{\sqrt{3}}{\hat x}_8$.

The form of the Hamiltonian components are dictated by the spin symmetry of the 
boundary Hamiltonian. We restrict here to the (dominant) 1- and 2-body terms of
$H_{\rm local}$. The unique one-body (diagonal) term can be written as 
a chemical potential term ${\cal H}_2$(r):
\begin{equation}
\sum_i {\hat x}_2^i = \frac{3}{\sqrt{2}}\sum_{i}  (n_i-2/3)=\frac{3}{\sqrt{2}}{\cal H}_2  \, ,
\label{Eq:mu}
\end{equation}
where $n_i=n_{i,0}+n_{i,1}$ counts the number of 0 or 1 on site $i$.
The diagonal 2-body density-density operators takes the form of a density-density (repulsive)
interaction ${\cal H}_{V}$:
\begin{eqnarray}
\sum_{i} {\hat x}_2^i{\hat x}_2^{i+r}&=&\frac{9}{2}\sum_{i} (n_i-2/3) (n_{i+r}-2/3)\nonumber \\
&=&\frac{9}{2}\,{\cal H}_{V}(r)\, ,
\label{Eq:nn}
\end{eqnarray}
Defining the pseudo-spin ${\mathbf S}=\frac{1}{2}\sum_{s,s'\in\{0,1\}} {\vec\sigma}_{ss'} |s\big>\big<s'|$
involving a combination of ${\hat x}_1$, ${\hat x}_3=\sqrt{3}|0\big>\big<1|$ and ${\hat x}_4=\sqrt{3}|1\big>\big<0|$, and combining 
three 2-body terms (that appears in $H_1$ and $H_{\rm local}$ with the same weights) we obtain an
effective Heisenberg-like couplings ${\cal H}_J(r)$:
\begin{eqnarray}
\sum_{i} ({\hat x}_1^i{\hat x}_1^{i+r}+{\hat x}_3^i{\hat x}_4^{i+r}+{\hat x}_4^i{\hat x}_3^{i+r})
&=&6\sum_{i}  {\bf S}_i\cdot {\bf S}_{i+r} \nonumber \\
&=& 6\,{\cal H}_J (r) \, .
\label{Eq:SS}
\end{eqnarray}
By symmetry, one also get (short-range) hopping terms ${\cal H}_t(r)$ by combining:
\begin{eqnarray}
\sum_{i} ({\hat x}_7^i{\hat x}_5^{i+r} \!\!&+&\!\!{\hat x}_5^i{\hat x}_7^{i+r} 
+ {\hat x}_8^i{\hat x}_6^{i+r} + {\hat x}_6^i{\hat x}_8^{i+r})\nonumber \\
&=&3\sum_{i,s} (b_{i+r,s}^\dagger b_{i,s} + b_{i,s}^\dagger b_{i+r,s})\nonumber\\&=&3\,{\cal H}_t(r) \, , 
\label{Eq:hopping}
\end{eqnarray}
and Josephson couplings ${\cal H}_\Delta(r)$ by combining:
\begin{eqnarray}
\sum_{i} ( {\hat x}_6^i{\hat x}_5^{i+r} \!\!&-&\!\! {\hat x}_5^i{\hat x}_6^{i+r} 
+ {\hat x}_8^i{\hat x}_7^{i+r} - {\hat x}_7^i{\hat x}_8^{i+r}) \nonumber\\
&=&3\sum_{i} (b_{i,0} b_{i+r,1} - b_{i,1}b_{i+r,0}) + h.c. \nonumber\\
&=&3{\cal H}_\Delta(r)\, ,
\label{Eq:pair}
\end{eqnarray}
which describe fluctuations of (s-wave) short-range singlet pairs.
The local Hamiltonian takes then the final form~:
\begin{eqnarray}
H_{\rm local}&=&c_0N_v+\frac{3c_2}{\sqrt{2}}\, {\cal H}_2+\sum_r V_r {\cal H}_V(r) \nonumber \\
&+&\sum_r t_r {\cal H}_t(r)+\sum_r  J_r{\cal H}_J(r)\nonumber \\
&+&\sum_r \Delta_r {\cal H}_\Delta(r)+H_{\rm rest}\, ,
\label{Eq:Hlocal}
\end{eqnarray}
where $H_{\rm rest}$ contains all negligible $N>3$ contributions.
The new physical parameters are simply related to the amplitudes appearing in the
expansion (\ref{Eq:Hb0}) of $H_{\rm local}$:
$t_r=3\, d_{57}(r)=3\, d_{68}(r)$, $J_r=6\, d_{11}(r)=6\, d_{34}(r)$, $V_r=\frac{9}{2}\, d_{22}$ and 
$\Delta_r=3\, d_{65}(r)=3\, d_{87}(r)$. 

For the $\mathbb{Z}_2$ RVB liquid on the kagome lattice, as seen on Fig.~\ref{Fig:two-body}(b),
all weights $d_{\lambda\mu}^2(r)$ (and hence all the
physical parameters $t_r$, $J_r$, $V_r$ and $\Delta_r$ decay exponentially fast with $r$  
so that $H_{\rm local}$ is a truly local operator.  
The dominant 2-body contribution to $H_{\rm local}$ is the (negative) 
hopping term. The density-density
interaction is attractive between nearest-neighbor sites ($V_1<0$) while it becomes repulsive 
(and very small) at longer distance ($V_r>0$ for $r\ge 2$). Finally, we note that the small 
Heisenberg spin interaction is {\it ferromagnetic} at all distances ($J_r<0$).
For the critical RVB wave function on the square lattice, as seen on Fig.~\ref{Fig:two-body}(a),
all weights $d_{\lambda\mu}^2(r)$ also decay exponentially fast with $r$ {\it except}
the (diagonal) density-density interaction ${\cal H}_V(r)$ which remains long-range. 
These remarkable features are to be connected to
the bulk correlations of the RVB wavefunctions: short-range (critical) bulk correlations 
translate into short-range (long-range) boundary Hamiltonians.
We have therefore established a one-to-one correspondence between the long-range behavior 
of the bulk correlations and the range of the boundary Hamiltonian of RVB wavefunctions.
This extends the previous findings~\cite{ciracpoilblanc2011} to the case of topological order.

\subsection{Entanglement spectra and edge modes} 

\begin{figure}\begin{center}
 \includegraphics[width=1.0\columnwidth]{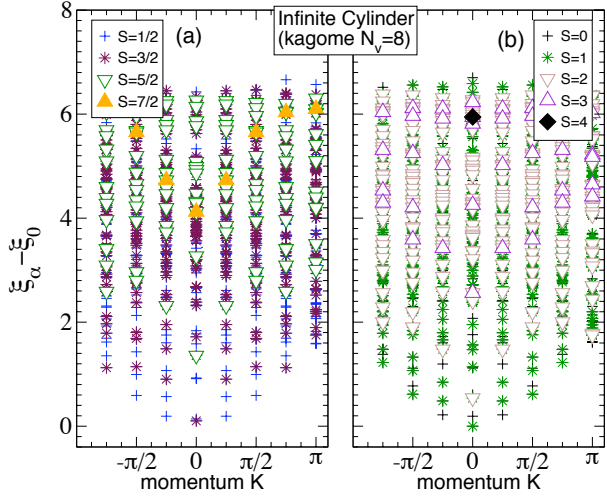}
 \end{center}
\caption{(Color online)
ES (w.r.t. the {\it same} $S_z=0$ GS energy $\xi_0$ at $K=0$) of an infinitely long kagome cylinder of perimeter $N_v=8$. 
Eigenstates with half-integer (a) and integer (b) spins correspond to odd and even sectors, respectively (see text).
}
\label{Fig:rvb2}
\end{figure}

\begin{figure}\begin{center}
 \includegraphics[width=1.0\columnwidth]{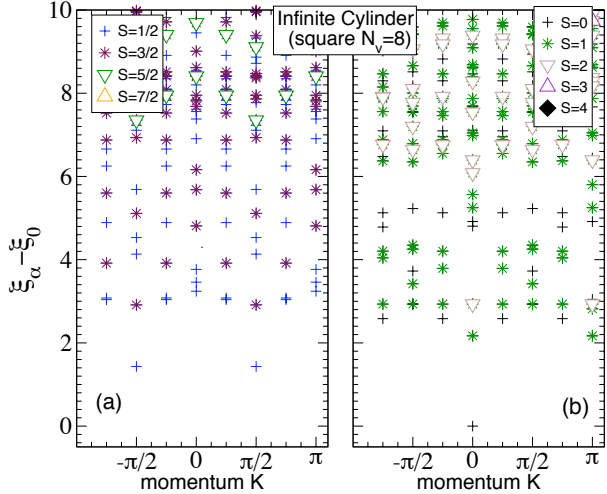}
 \end{center}
\caption{(Color online)
Same as Fig.~\protect\ref{Fig:rvb2} for the critical RVB state (square lattice). 
Eigenstates with half-integer (integer) spins correspond to $\Delta$ odd (even) -- see text.
For OBC one gets a subset of (b) ($\Delta=0$ sector).
}
\label{Fig:rvb0}
\end{figure}

We now move to the investigation of the full bipartite ES which is given by the spectrum of $H_{\rm local}$.
Our results are summarized in Figs.~\ref{Fig:rvb2}(a,b) and \ref{Fig:rvb0}(a,b), for
infinitely-long cylinders with kagome and square lattices. 
For convenience, the GS energy of $H_{\rm local}$ (corresponding to the largest weight in the RDM)
is subtracted from the spectra.
The (excitation) ES are shown as a function of momentum around the cylinder
and the eigenstates are labelled according to their spin-multiplet structure
inherited from the su(2) symmetry of the RVB state,
although with the $1/2\oplus 0$ representation.~\cite{perez_2010} 
A careful finite size scaling (see Appendix B) suggests
that the kagome (square) lattice cylinder ES is gapless (gapped) in the limit
$N_v\rightarrow\infty$. Since these features are opposite to what is expected
for the energy excitation spectra of the corresponding bulk systems (according to their
long wavelength properties), we deduce that the ES characterizes specifically 
the nature of the L and R edge modes (Fig.~\ref{Fig:cylinders}).  
Note that for given choice of BC, the actual ES is the spectrum  
of a {\it projected} $H_1$ Hamiltonian and, hence, is a subset of the full ES.
For example, in a kagome lattice (square lattice) cylinder with OBC, a common set-up in numerical simulations,
only (a sub-set of) the integer spin eigenstates are obtained. 


\subsection{Topological entropy}

It is of great interest to investigate the entanglement entropy which can give access to the quantum dimension $\cal D$ and hence provides clear fingerprints of topological order~\cite{wen-91}. 
We recall that we consider here the RVB wavefunction for which the {\it same} sign enters in the linear
superposition of the dimer coverings (defined, on the kagome lattice, as $\Psi_{\rm RVB}^+$ with no vison line,
i.e. no $\mathbb{Z}_2$ flux through the cylinder).
We investigate infinite cylinders and study the behavior of the EE as a function of the perimeter.
From specific choices of the cylinder boundary conditions we can select specific conserved sectors on the edge (e.g. 
OBC for the kagome cylinder selects the even sector, etc...). 
The EE is given by the Von Neumann entropy $S_{VN}=-{\rm Tr} \{\sigma_b^2 \ln{\sigma_b^2}\}$.

\begin{figure}\begin{center}
 \includegraphics[width=1.0\columnwidth]{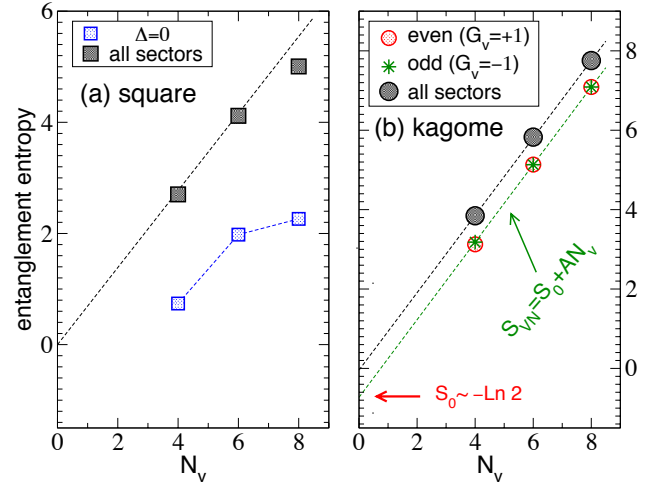}
 \end{center}
\caption{(Color online)
Entanglement entropy versus perimeter 
$N_v$ for specific sectors (open symbols) or when summing over all sectors (shaded symbols). 
(a) square lattice ($\Delta=0$ is obtained with OBC); (b) kagome lattice 
(no $\mathbb{Z}_2$ flux through the cylinder, $G_h=0$)}
\label{Fig:SVN}
\end{figure}

As shown in Fig.~\ref{Fig:SVN}(a), the EE of the square lattice cylinder
with OBC ($\Delta=0$ sector) shows strong deviations from the area law
(i.e. linear behavior with $N_v$) which should be connected to the
critical nature of the RVB wave function.  In contrast, for both even and
odd (edge) sectors of the kagome RVB wavefunction, the EE can be well
fitted according to $S_{VN}=S_0 + A N_v$, where $S_0=-\ln{\cal D}$ is the
topological EE, as shown in Fig.~\ref{Fig:SVN}(b). The existence of a
finite $S_0$, a smoking gun of the topological nature of the RVB state,
can be seen as a direct consequence of the particular structure of $H_1$
according to the following argument: The EE is given (crudely) by
$-\ln{\cal N}$ where $\cal N$ is the number of eigenstates of $H_1$ below
a fixed energy scale of order 1.  For fixed cylinder boundaries, the
support of $H_1=H_{\rm local}{\cal P}$ (${\cal P}={\cal P}_{\rm even}$ or
${\cal P}={\cal P}_{\rm odd}$) contains ${\cal N}\simeq\frac{1}{2}\,
3^{Nv}$ states and $S_0=-\ln{2}$, as expected for a topological
$\mathbb{Z}_2$ spin liquid with quantum dimension ${\cal D}=2$.  Note that
$\Psi_{\rm RVB}^{+,\mathrm{even}}$ and $\Psi_{\rm RVB}^{+,\mathrm{odd}}$
can be seen as ``minimally entangled states" naturally produced by the
DMRG algorithm from amongst the quasi-degenerate ground states of the
$\mathbb{Z}_2$ topological phase~\cite{Balents2}. Linear combination of
them (or e.g. of $\Psi_{\rm RVB}^{+,\mathrm{even}}$ and $\Psi_{\rm
RVB}^{-,\mathrm{even}}$) should give a larger topological entropy.

Summing over all sectors amounts to taking $H_1=H_{\rm local}$ so that all
eigenstates of the ES contribute and $S_{VN}\propto N_v$ (as can be shown
rigorously) as seen in Fig.~\ref{Fig:SVN}(b).  For the square lattice,
severe constraints leads to an {\it extensive} number 
(i.e. proportional to the perimeter $N_v$) of disconnected sectors on the edge of
dimension ${\cal N}\simeq\frac{1}{N_v}\, 3^{Nv}$, therefore introducing {\it negative} logarithmic corrections 
$\sim -\ln{N_v}$ to the EE for any boundary conditions (see e.g. data for OBC on Fig.~\ref{Fig:SVN}(a)).
The long-range diagonal interaction in $H_{\rm local}$ (Fig.~\ref{Fig:two-body}(a)) may also be
responsible for deviations from the area law, even when considering all sectors.

\section{Discussion and outlook}

Introducing PEPS representations and using Tensor Networks techniques, 
we have examined topological and entanglement properties associated to gapped and gapless RVB states
using cylindrical geometries with arbitrary boundary conditions. The formalism allows to take the limit of infinite cylinders. 
Using the simple topological structure of the space of dimer coverings on the kagome lattice, we construct four quasi-degenerate (for a generic quantum HAF) orthogonal RVB states and  obtain the finite size scalings of the energy splittings amongst them (topological gaps) which could be compared 
to numerical simulations. Incidentally, our results identify two very different energy splittings decaying 
with two clearly different length scales. The largest energy scale corresponds to inserting a (horizontal) vison line (or a $\mathbb{Z}_2$ flux in the cylinder).
The second energy scale corresponds to pinning a site with a strong field
on either end of the cylinder, which would be the same as moving a site from one end to the other.  
Although it has been suggested that the DMRG algorithm (naturally) selects a minimally entangled state~\cite{Balents2},
it is still not clear how to reconcile the fact 
that the finite size corrections of the groundstate energy are very small~\cite{White,Schollwoeck} 
while our RVB computation predicts clear finite-size effects for the states with a definite 
 $\mathbb{Z}_2$ flux. 
On the other hand, we find that the energy {\it averaged} 
over the four (minimally entangled) RVB states shows very small finite size effects.

In addition, we show that boundary Hamiltonians can be written as $H_{\rm local}{\cal P}+
\beta_{\rm topo} ({\rm Id}-{\cal P})$, $\beta_{\rm topo}\rightarrow\infty$. 
In particular, we have established the existence of
a projector $\cal P$ (which intrinsically depends on the boundary conditions) onto a restricted 
subspace at the edge (as for Kitaev toric 
code~\cite{kitaev:toriccode,ciracpoilblanc2011}), a consequence of the disconnected 
topological sectors in the space of dimer coverings of the lattice.
We argue that the non-local character of the resulting boundary Hamiltonian
is the fingerprint of topology. The ES is a subset (associated to $\cal P$) of the 
spectrum of the emerging 
local Hamiltonian $H_{\rm local}$ acting on the unrestricted edge space. 
In contrast to the toric code for which $H_{\rm local}$ is trivial,
here $H_{\rm local}$ takes the form of a short-range (bosonic) t--J 
model (including a long-range diagonal interaction for the critical RVB state). 
We argue that the topological features (e.g. finite size scaling of topological gaps) and entanglement properties (e.g.
structure of boundary Hamiltonians) of the NN RVB wave functions are characteristic of topological phases.
We propose to use these features to detect topological order
in microscopic models.~\cite{Assaad,White,Balents,Balents2,Schollwoeck}

\section*{ACKNOWLEDGEMENT}

D.P. acknowledges support by the ``Agence Nationale de la Recherche" under grant No.~ANR~2010~BLANC~0406-0.  This work was granted access to the HPC resources of CALMIP under the allocation 2012-P1231. 
D.P. also thanks Steve R. White for sharing DMRG results and Steve A. Kivelson for 
useful correspondence. 
N.S. acknowledges
helpful discussions with Frank Verstraete and
support by the Alexander von Humboldt foundation,
the Institute for Quantum Information and
Matter (an NSF Physics Frontiers Center with support of
the Gordon and Betty Moore Foundation) and the NSF
Grant No. PHY-0803371.
D.P.-G. acknowledges QUEVADIS and Spanish grants QUITEMAD and MTM2011-26912.
J.I.C. acknowledges the EC project Quevadis, the DFG Forschergruppe 635, and Caixa Manresa.
This work was initiated at Centro de Ciencias Pedro Pascual (Benasque, Spain). 

\section*{APPENDIX A: TAKING THE LIMIT OF THE INFINITE CYLINDER}

\begin{figure}\begin{center}
 \includegraphics[width=0.8\columnwidth]{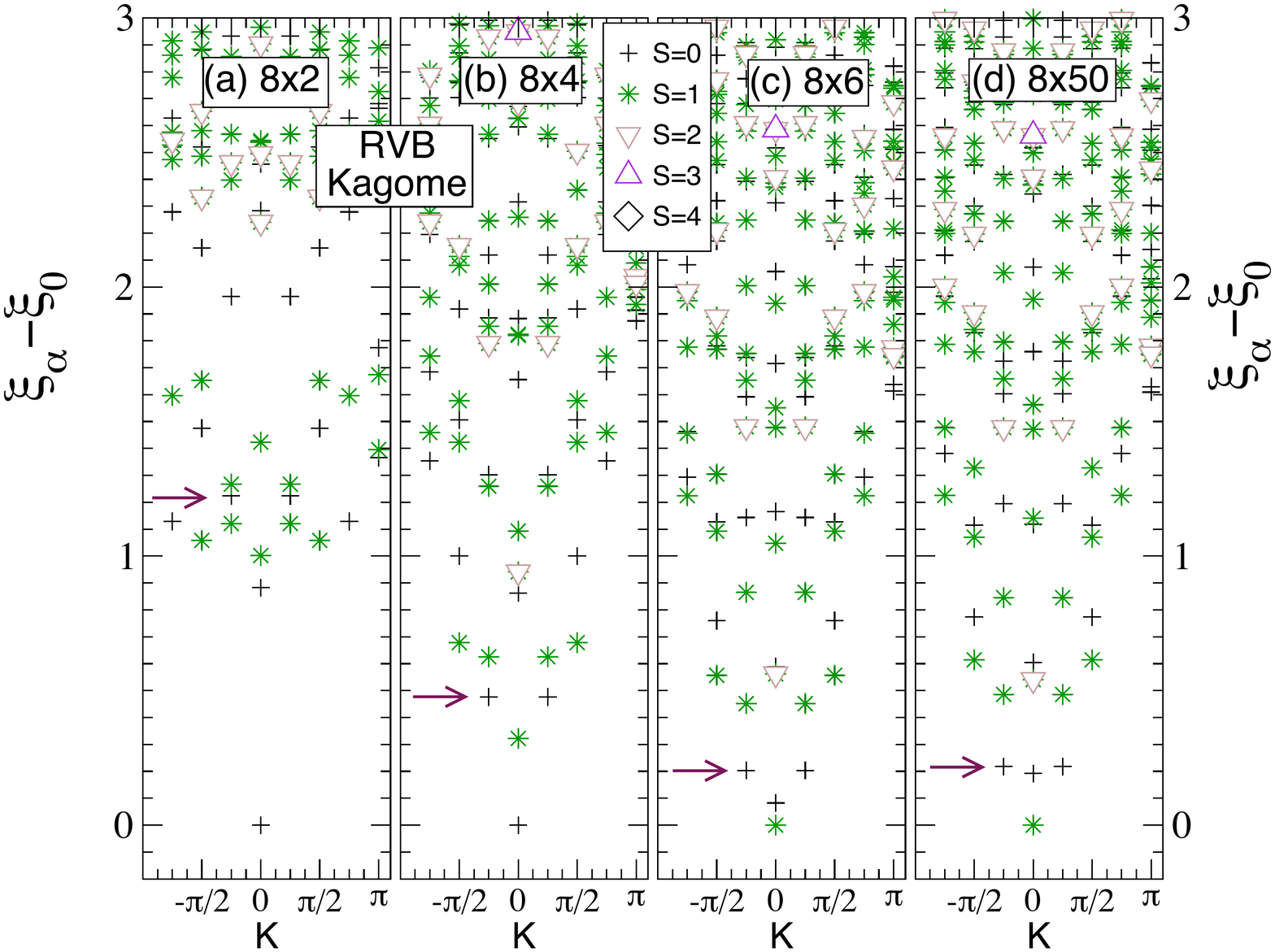}
 \includegraphics[width=0.8\columnwidth]{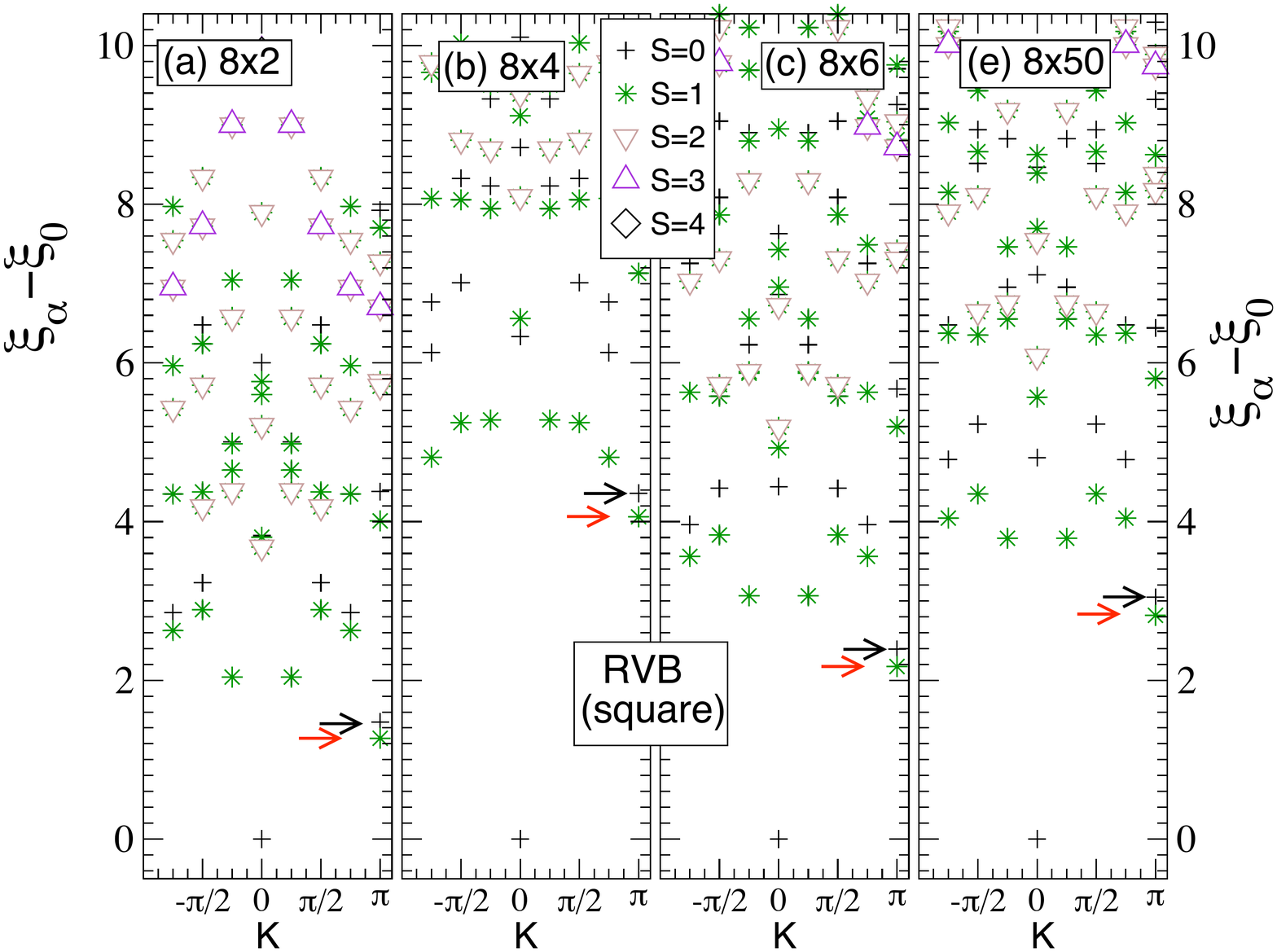} \end{center}
\caption{(Color online)
ES of a cylinder with fixed perimeter $N_v=8$ and increasing length $N_h$ ranging from
$2$ to $50$. OBC are used for $B_L$ and $B_R$. Kagome (top) and square (bottom) lattices.}
\label{Fig:ladders3}
\end{figure}

\begin{figure}
\begin{center}
 \includegraphics[width=0.9\columnwidth]{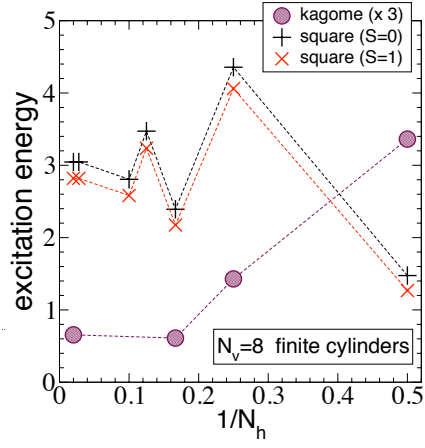}
 \end{center}
\caption{(Color online)
Finite size scaling of the lowest energy levels marked by arrows in 
the ES of Fig.~\ref{Fig:ladders3}. 
Excellent convergence is found when $N_h\rightarrow\infty$ (at constant $N_v=8$).
Note the alternating behavior according to the parity of $N_h/2$ for the square lattice, in contrast to the kagome 
lattice showing a straight exponential convergence.
}
\label{Fig:gaps_vsNh}
\end{figure}

The (excitation) ES are shown in Fig.~\ref{Fig:ladders3}, 
as a function of momentum around the cylinder
and the eigenstates are labelled according to their spin-multiplet structure
inherited from the su(2) symmetry of the RVB state,
although with the $1/2\oplus 0$ representation.
From the data shown in Fig.~\ref{Fig:ladders3}, we see
that,  for a fixed perimeter, the ES converge rapidly when increasing the length towards 
the infinite-cylinder limit ($N_h=\infty$). The latter is reached as soon as
$N_h>N_v$ ($N_h\gg N_v$) for the Kagome (square) lattice: 
in practice, the RDM for $N_h=50$ is fully converged. 
This is clear from the finite size scaling analysis of some of the low-energy excitations of the 
ES shown in Fig.~\ref{Fig:gaps_vsNh}. 

\section*{APPENDIX B: FINITE SIZE SCALING OF THE ENTANGLEMENT SPECTRUM VS CYLINDER PERIMETER}

\begin{figure}\begin{center}
 \includegraphics[width=0.8\columnwidth]{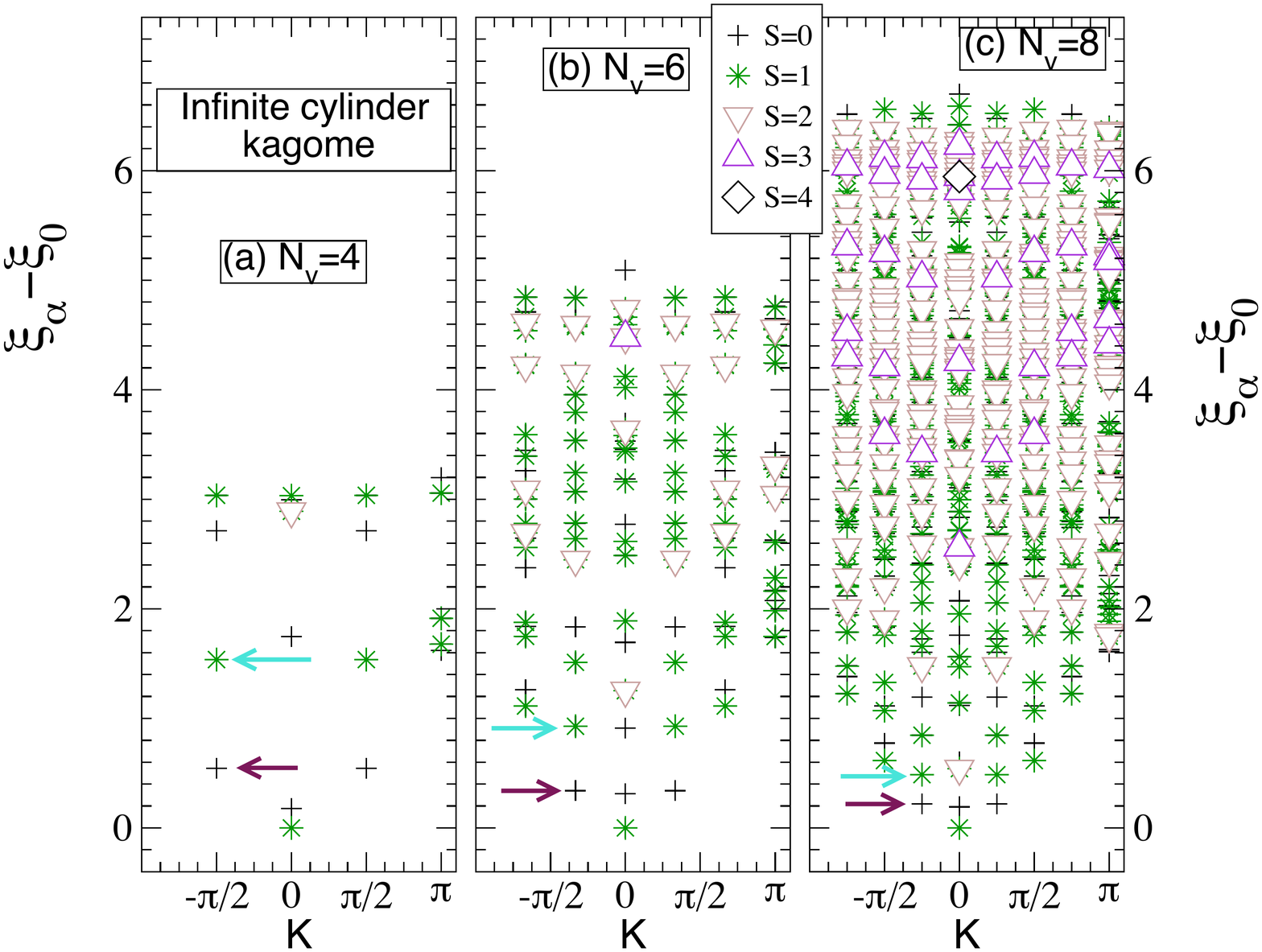}
 \includegraphics[width=0.8\columnwidth]{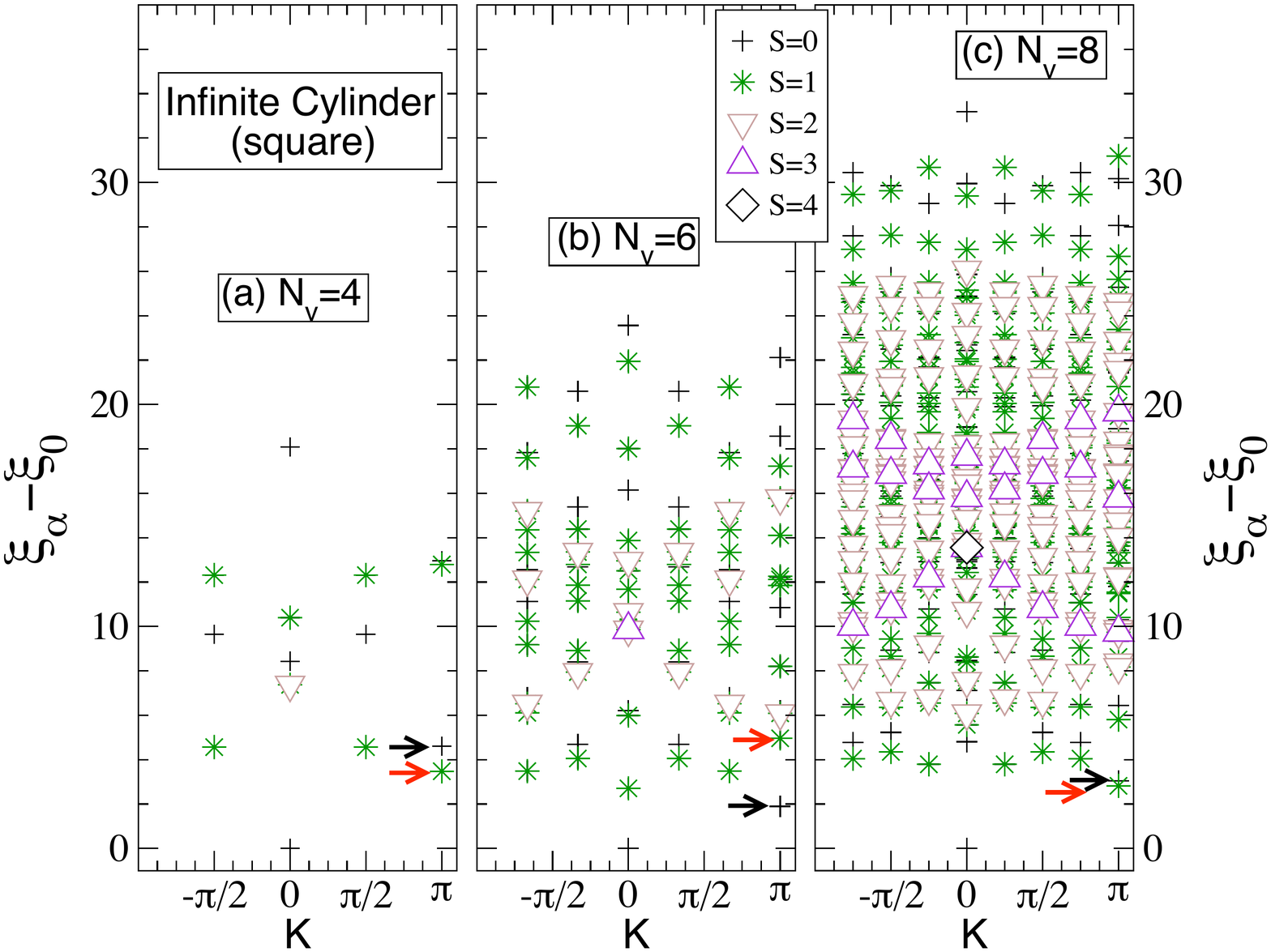}
 \end{center}
\caption{(Color online)
ES of infinite cylinders with increasing perimeter $N_v$.
OBC are used for $B_L$ and $B_R$. Kagome (top) and square (bottom) lattices.}
\label{Fig:cyl2}
\end{figure}

\begin{figure}\begin{center}
 \includegraphics[width=0.9\columnwidth]{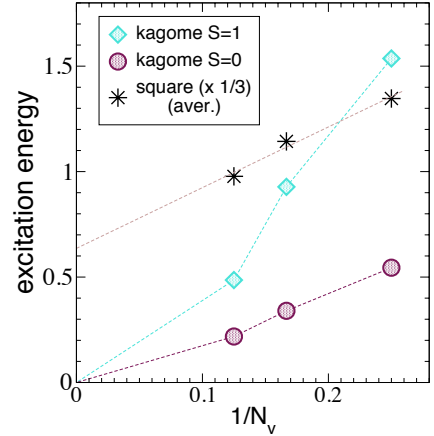}
 \end{center}
\caption{(Color online)
Finite size scaling of low excitation energies of 
 infinite cylinders vs inverse perimeter, suggesting a vanishing (finite) gap in the thermodynamic limit
for the kagome (square) lattice. For the square lattice, the average (divided by 3) 
between the lowest $K=\pi$ singlet and triplet excitations is shown.
}
\label{Fig:gaps_vsNv}
\end{figure}

After taking the limit of the infinite cylinder ($N_h\rightarrow\infty$), we investigate
the dependance of the ES as a function of the cylinder perimeter $N_v$.
Note that when taking the $N_h\rightarrow\infty$ limit, one still has to specify the choice of
the $B_L$ and $B_R$ boundaries that uniquely determine the projector (even or odd) involved at the edges along the cut.
Here we only consider OBC which select the integer spin sector of the boundary Hamiltonian.
Our results are summarized in Fig.~\ref{Fig:cyl2}, for
infinitely-long cylinders with kagome and square lattices. 
A careful analysis of these spectra and of some of their low-energy excitations
(see Fig.~\ref{Fig:gaps_vsNv}(a)) as a function of cylinder perimeter suggests
that the square (kagome) lattice cylinder ES is gapped (gapless) in the limit
$N_v\rightarrow\infty$. Since these features are opposite to what is expected
in the corresponding bulk systems, we 
deduce that the ES characterizes specifically 
the nature of the edges (L and R in Fig.~\ref{Fig:cylinders}(a-d)).  

\section*{APPENDIX C: EXPANSION IN TERMS OF MANY-BODY OPERATORS}

Any operator ${\cal O}_{\rm edge}$ like projectors ${\cal P}$ or boundary Hamiltonians 
acting on the edge can be expanded in terms of $3^{2 N_v}$
orthogonal (real) operators $\hat X_\alpha$,
\begin{equation}
{\cal O}_{\rm edge}=c_0 N_v  +\sum_{\alpha} A_\alpha  \, \hat X_\alpha \, .
\label{Eq:Hb}
\end{equation}
The scalar product 
in the operator basis is defined as $\big< {\hat u}{\hat  v} \big>$, 
where $\big< \cdots\big>=\frac{1}{Z}{\rm Tr}(\cdots)$ and the trace is {\it a priori} performed over the
full basis of $Z=3^{N_v}$ states. For convenience, the constant term $c_0=\frac{1}{N_v}\big< {\cal O}_{\rm edge}\big>$
has been separated so that we can assume all other operators
satisfy ${\rm Tr} \, \hat X_\alpha =0$.
Simple algebra shows that the coefficients can be obtained by taking the trace
of the corresponding operators with ${\cal O}_{\rm edge}$ as,
\begin{equation}
A_\alpha= \big< \hat X_\alpha {\cal O}_{\rm edge} \big> \,   /   \big< \hat X_\alpha \hat X_\alpha^\dagger\big>
\, ,
\label{Eq:Amplitude}
\end{equation}
where the trace in the numerator involves, in fact, the sum over the projected subspace. 
We also obtain some "sum-rule":
\begin{equation}
\big< {\cal O}_{\rm edge}^{\, 2}\big> =(c_0 N_v)^2   +\sum_\alpha   A_\alpha^2 \big< \hat X_\alpha 
\hat X_\alpha^\dagger\big>\, ,
\label{Eq:SumRule}
\end{equation}
which enable to compute the weight associated to each operator. 

To go further and expand ${\cal O}_{\rm edge}$ in the full operator basis, 
it is convenient to use a local basis of 9 (normalized)
operators $\{ {\hat x}_0,\cdots, {\hat x}_8\}$ which act 
on the local site configuration, $\{ |0\big>, |1\big>, |2\big> \}$, e.g.
${\hat x}_0={\bold 1}$, ${\hat x}_1=\sqrt{\frac{3}{2}}(|0\big>\big<0| -|1\big>\big<1|)$ 
and ${\hat x}_2=\frac{1}{\sqrt{2}}(|0\big>\big<0|+|1\big>\big<1|-2|2\big>\big<2|)$,
for the diagonal matrices, 
complemented by  
$\hat x_3=\hat x_4^\dagger=\sqrt{3}|0\big>\big<1|$
acting as ``spin" operators,
and $\hat x_5=\hat x_7^\dagger=\sqrt{3} |2\big>\big<0|$ and 
$\hat x_6=\hat x_8^\dagger=\sqrt{3} |2\big>\big<1|$
acting as annihilation and creation (hardcore) bosonic operators.
These operators
satisfy ${\rm tr}({\hat x}_\lambda)=0$ (for $\lambda\ne 0$)
and ${\rm tr}({\hat x}_\lambda {\hat x}_\lambda^\dagger)=3$,
where ``tr" is the trace over the local degrees of freedom (of some site $i$).
From now on, we extend the action of these local operators to the whole edge,
assuming a trivial (implicit) action on the $N_v-1$ unspecified sites, i.e. 
${\hat x}_\lambda^i \equiv
{\hat x}_\lambda^i \otimes {\bf 1}^{\otimes (N_v-1)}$, so that  
${\rm Tr} ({\hat x}_\lambda^i ({\hat x}_\lambda^i)^\dagger)=3^{N_v}$
and $\big< {\hat x}_\lambda^i ({\hat x}_\lambda^i)^\dagger\big>=1$.
Using the local basis of operators, one can then uniquely expand any edge operator like
$H_1$ in terms of N-body operators
as,
\begin{eqnarray}
{\cal O}_{\rm edge} &=& c_0 N_v+\sum_\lambda c_{\lambda} \sum_i{\hat x}_\lambda^i 
+ \sum_{\lambda,\mu,r} d_{\lambda\mu}(r) 
\sum_{i}^\prime  {\hat x}_\lambda^i {\hat x}_\mu^{i+r} \nonumber \\
&+& \sum_{\lambda,\mu,\nu,r,r'} e_{\lambda\mu\nu}(r,r') 
\sum_{i}^\prime {\hat x}_\lambda^i {\hat x}_\mu^{i+r} {\hat x}_\nu^{i+r'} 
+ \cdots \, ,\label{Eq:Hb_bis}
\end{eqnarray}
where each group of terms involves products of $N=1,2,\cdots,N_v$ on-site operators
${\hat x}_\lambda^i$, $i$ labeling the sites.
Here the sums do not contain the identity, the sums over distances are restricted to non-equivalent
relative distances.
$\sum^\prime$ means that only translations giving {\it distinct} sets of sites are performed (no 
multiple counting). Hence,
the N-body {\it translationally invariant} operators $\hat X_\alpha$ in (\ref{Eq:Hb_bis}),
where $\alpha$ combines all the labels of the coefficients of the expansion
(e.g. $\alpha=(\lambda,\mu,r)$ and 
$\hat X_\alpha=\sum_{i}^\prime  {\hat x}_\lambda^i {\hat x}_\mu^{i+r}$), 
are normalized as $\big<\hat X_\alpha \hat X_\alpha^\dagger\big>=N_v/g_\alpha$,
where $g_\alpha$ are ``multiplicity" factors that count the number
of times the operator maps onto itself under all $N_v$ translations.
The (real) coefficients in (\ref{Eq:Hb_bis}) are obtained by taking the trace (in operator space)
of the corresponding operators with the operator ${\cal O}_{\rm edge}$,
\begin{eqnarray}
c_{\lambda} &=&  \frac{1}{N_v} \big<  (\sum_{i=1}^{N_v} {\hat x}_\lambda^i) {\cal O}_{\rm edge} \big> \, ,\\
d_{\lambda\mu}(r) &=&  \frac{1}{N_v} \big<(\sum_{i=1}^{N_v} {\hat x}_\lambda^i {\hat x}_\mu^{i+r}) {\cal O}_{\rm edge}  \big> \ ,\\
e_{\lambda\mu\nu}(r,r') &=&  \frac{1}{N_v} \big< (\sum_{i=1}^{N_v}{\hat x}_\lambda^i {\hat x}_\mu^{i+r}) {\hat x}_\nu^{i+r'} 
{\cal O}_{\rm edge} \big>  \, ,
\end{eqnarray}
where one can make advantage of translation symmetry to compute the r.h.s of these equations. 
The sum-rule for the weights takes then the form: 
\begin{eqnarray}
\frac{1}{N_v}\big< ({\cal O}_{\rm edge})^2\big> &=& c_0^{\, 2} N_v+\sum_\lambda c_{\lambda}^{\, 2} 
+ \sum_{\lambda,\mu} \sum_r \frac{1}{g_r} d^{\, 2}_{\lambda\mu}(r) \nonumber \\
&+& \sum_{\lambda,\mu,\nu}\sum_{r,r'} \frac{1}{g_{r,r'}} e^{\, 2}_{\lambda\mu\nu}(r,r') +\cdots 
\label{Eq:SumRule-bis}
\end{eqnarray}
where the (second) 
sums are restricted to non-equivalent sets of distances and the multiplicity factors only
depend on the latter.


\begin{thebibliography}{99}
\bibitem{note1} Nevertheless, a dimer wavefunction on the Kagome lattice can be mapped to Kitaev's toric code on the square lattice, which indicates that in the former case there should exist a non-local boundary Hamiltonian.~\cite{long_paper}
\bibitem{note_CL} The critical or short-range behavior of the RVB wavefunctions can also be inferred from 
finite size analysis of the correlation lengths obtained from the PEPS transfer matrices.
\bibitem{note2}  On the kagome lattice, $H_{\rm local}$ corresponding to the L and R edges
differ slightly. Here, we refer to the L edge.
\end{thebibliography}
\end{document}